\documentclass[fleqn,usenatbib]{mnras}
\hypersetup{draft}

\usepackage{newtxtext,newtxmath}

\usepackage{float}
\usepackage[caption = false]{subfig}

\usepackage[T1]{fontenc}
\usepackage{ae,aecompl}

\usepackage{graphicx}	
\usepackage{amsmath}	
\usepackage{amssymb}	
\usepackage{threeparttable}
\usepackage{longtable}

\newcommand{\EQ}[1] {Equation~(\ref{#1})}
\newcommand{\SEC}[1] {Section~\ref{#1}}
\newcommand{\APP}[1] {Appendix~\ref{#1}}
\newcommand{\FIG}[1] {Figure~\ref{#1}}
\newcommand{\TAB}[1] {Table~\ref{#1}}

\newcommand{\BEA}{\textsc{bear}}

\newcommand{\VEC}[1] {{\boldsymbol{{ #1}}}}

\title[FRB searching using NS26m and KM40m]{Piggyback 
search for fast radio bursts using Nanshan 26m and Kunming 40m radio 
telescopes -- I. Observing and data analysis systems, discovery of a mysterious 
peryton}

\author[Y. P. Men et al.]
{Y. P. Men$^{1, 2}$\thanks{E-mail: ypmen@pku.edu.cn},
R. Luo$^{1, 2}$,
M. Z. Chen$^{4}$,
L. F. Hao$^{5}$,
K. J. Lee$^{1, 3}$\thanks{E-mail: kjlee@pku.edu.cn},
J. Li$^{4}$,
Z. X. Li$^{5}$,\newauthor
Z. Y. Liu$^{4}$,
X. Pei$^{4}$,
Z. G. Wen$^{4}$,
J. J. Wu$^{6}$,
Y. H. Xu$^{5}$,
R. X. Xu$^{1, 2}$,
J. P. Yuan$^{4}$,\newauthor
C. F. Zhang$^{1, 2}$
\\
$^{1}$Kavli Institute for Astronomy and Astrophysics, Peking University, 
Beijing 100871, P.R. China\\
$^{2}$Department of Astronomy, School of Physics, Peking University, Beijing 100871, China\\
$^{3}$National astronomical observatory, CAS, Beijing, 100012, China\\
$^{4}$Xinjiang Astronomical Observatory, CAS, 150 Science 1-Street, Urumqi, Xinjiang 830011, China\\
$^{5}$Yunnan Astronomical Observatory, Chinese Academy of Sciences, Kunming 650011, China\\
$^{6}$School of Electronics Engineering and Computer Science, Peking University, Beijing 100871, China
}

\date{Accepted 2019 July 02. Received 2019 July 02; in original form 2019 February 08}

\pubyear{2019}
\begin{document}
\label{firstpage}
\pagerange{\pageref{firstpage}--\pageref{lastpage}}
\maketitle

\begin{abstract}
We present our piggyback search for fast radio bursts using the Nanshan 26m Radio Telescope and the Kunming 40m Radio Telescope. The observations are performed  in the L-band from 1380 MHz to 1700 MHz at Nanshan and S-band from 2170 MHz to 2310 MHz at Kunming. We built the \textsc{Roach2}-based FFT spectrometer and developed the real-time transient search software. We introduce a new radio interference mitigation technique named \emph{zero-DM matched filter} and give the formula of the signal-to-noise ratio loss in the transient search. Though we have no positive detection of bursts in about 1600 and 2400 hours data at Nanshan and Kunming respectively, an intriguing peryton was detected at Nanshan, from which hundreds of bursts were recorded. Perytons are terrestrial radio signals that mimic celestial fast radio bursts. They were first reported at Parkes and identified as microwave oven interferences later. The bursts detected at Nanshan show similar frequency swept emission and have double-peaked profiles. They appeared in different sky regions in about tens of minutes observations and the dispersion measure index is not exactly 2, which indicates the terrestrial origin.  The peryton differs drastically from the known perytons detected at Parkes, because it appeared in a precise period of $p=1.71287\pm 0.00004$ s. Its origin remains unknown.
\end{abstract}




\section{Introduction}

\citet{Lorimer2007} discovered the first fast radio burst (FRB). 
At first, it was unclear whether this was a new type of celestial
radio source, or radio frequency interference (RFI).
Particularly, FRBs share two major similarities with the ground-based RFI:
1) both have very bright flux ($\sim$0.3 Jy to 
$\sim$100 Jy); 
2) both have durations of a few milliseconds. 
Furthermore, since the Lorimer 
burst was not observed again in the follow-up observations, 
it was impossible to assert the source origin. 

On the other hand, Lorimer burst may well be celestial. The burst signal showed 
a characteristic cold plasma dispersion relation, that the group delay at 
frequency $\nu$ is \begin{equation}
	t = 4.149\,{\rm DM}\left(\nu_{\rm 1,GHz}^{-2}-\nu_{\rm 2,GHz}^{-2}\right)\,{\rm ms}\,.
	\label{eq:dispersion}
\end{equation}
The dispersion measure, DM, 
is the column density of free electrons in the unit of pc~cm$^{-3}$
along the line of sight, i.e.  
${\rm DM}\equiv\int{n_\mathrm{e}dl}$ where $n_{\rm e}$ is the electron density.  
Due to the free electrons in the interstellar medium, such a cold plasma 
dispersion is observed extensively in pulsar signals \citep{Manchester81aj}.
The cold plasma dispersion seen in FRBs highly suggests
that they are of extraterrestrial origin.

Further investigation shows that the RFI may mimic the dispersive signatures.  
Such RFI signals are called \emph{perytons} and were first discovered at Parkes 
\citep{Burke_Spolaor11apj}. It was suggested \citep{Petroff2016} that Perytons 
differ from FRBs in the following properties:
1) Perytons are strongly clustered in DM and time of day, whereas FRBs are not.
2) FRBs follow the cold plasma dispersion, where some of the perytons show 
deviations from this relation;
3) FRBs, as far-field sources, are well localised on the sky. Perytons, 
if being near-field interference signals, appear to have multiple locations;
4) Perytons have longer average pulse durations than FRBs, e.g. the pulse 
durations of perytons concentrate around 30-40 ms while most FRBs have durations of 
a few milliseconds. Using a RFI monitor, \citet{Petroff15} identified the 
perytons as microwave oven interferences.
\citet{Kocz2012} reported that some perytons occurred approximately 22 s apart 
.

The celestial origin of FRB was confirmed by other discoveries.  Shortly after 
Lorimer's work, a growing number of FRBs were
discovered with Parkes at 1.4 GHz, either in archival data
\citep{Keane12MN,Thornton13Sci, Burke-Spolaor14ApJ} or from
real-time searches \citep{Ravi15ApJ,Petroff15MN, Keane16Nat,
Ravi16Sci, Petroff17MN, Bhandari18MN}. Telescopes other than
Parkes have also detected FRBs \citep{Spitler14ApJ, Masui15Nat,
Bannister17ApJ}, including interferometers \citep{Caleb17MN}. 
For more information on currently known FRBs ($\sim$ 65 of them), one can look up 
the online database\footnote{FRBCAT: \url{http://frbcat.org}}  by 
\citet{Petroff2016}. There are to-date two known repeating sources, namely FRB\,121102 and 
FRB\,180814, discovered by Arecibo and CHIME, respectively \citep{Spitler14ApJ, 
Spitler16, Amiri2019nat}. The FRB celestial origin is ultimately established with 
interferometry location and host galaxy discoveries. The host galaxy of 
FRB\,121102 was identified as a low-metallicity, star-forming dwarf galaxy and a 
persistent radio counterpart was found \citep{Chatterjee2017nat, 
Tendulkar2017apj, Marcote2017apj}.

In this paper, we report our realtime transient search project carried out at 
two Chinese telescopes. Our observations and discovery of an intriguing type of 
RFI are also presented. Unlike the previously reported perytons, 
the newly discovered RFI
shares more similarities with the FRBs, and it is probably not created by microwave ovens.
The paper is organised as follows.  In \SEC{sec:obsall}, we describe our 
piggyback observing scheme (\SEC{sec:obssch}),  hardware (\SEC{sec:hard}), and 
data reduction pipeline (\SEC{sec:rtpip}). We also investigate the sensitivity 
and event rate of our FRB searching in \SEC{sec:eventrate}. The interesting 
peryton is reported in \SEC{sec:peryton}. The related discussions and 
conclusions are made in \SEC{sec:dis}.

\section{Observing and realtime search pipelines}
\label{sec:obsall}
\subsection{Observations}
\label{sec:obssch}
We carried out observations with the Nanshan 26-metre radio telescope (NS26m) of 
\emph{Xinjiang Astronomical Observatory} (XAO) and the Kunming 40-metre radio 
telescope (KM40m) of \emph{Yunnan Astronomical Observatory} (YNAO).
NS26m is located 70 km away from the city of Urumqi. The position of the NS26m is
longitude E87$^{\circ}10'41''$, latitude N+43$^{\circ}29'17''$, and altitude
2080m. It was built in 1993 and had served as the general purpose 
centimeter-wavelength radio telescope for the Chinese astronomy community for 
more than 20 years \citep{WMZ01}. At 1.4 GHz (L-band), the radio 
frequency bandwidth is 320~MHz, and the system temperature is 23 K. 

KM40m was built in 2006 for the Chinese lunar-probe mission and gradually 
started performing scientific observations \citep{Hao2010}.  
It is located in the 
south west of China (${\rm N} 25^\circ01'38''$, ${\rm
E} 102^\circ47'45''$, altitude 1960m), approximately 15 kilometers away from the
nearby city of Kunming. There is a room-temperature S(2.2 GHz)/X(8.5 GHz) dual-band,
circularly-polarised receiver installed for satellite tracking
purposes. The X-band receiver is used in lunar-probe mission and its temperature is too high for our purposes, 
so we search in the S-band. The system temperature of the receiver at S band is 70 K at 2.2 GHz.
The radio frequency signal is down converted to the intermediate
frequency, which has 300 MHz bandwidth. However, due to RFI only about
140~MHz of clean bandwidth is available. The parameters of the telescopes are 
summarized in \TAB{tab:teles}. On average per week, we took 48-hour and 120-hour 
observations at NS26m and KM40m, respectively. The total amount of the raw data 
would be 30 TB per month, if all data were recorded.

\begin{figure}
	\includegraphics[width=\columnwidth]{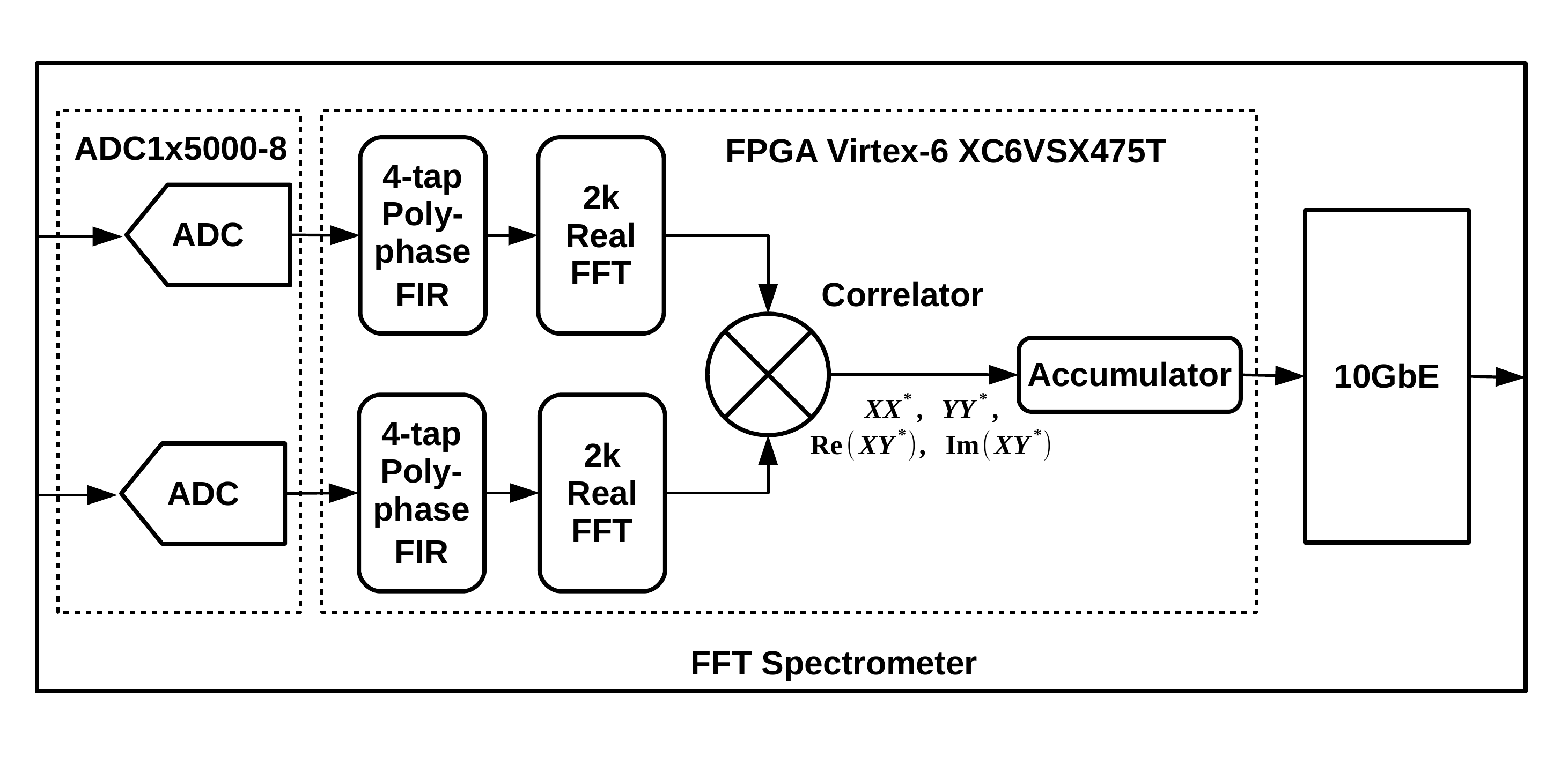}
		\caption{The spectrometer flowchart. The FX engine is implemented on 
		FPGA chip, which use the polyphase filterbank to channelize the data and 
		then do multiplication. The spectra is integrated and packetized into UDP 
		packets.  Each packet is then transmitted to the computer node using 10GbE 
		fiber links.}
    \label{fig:FX_correlator}
\end{figure}

We note that most of the RFI signals at KM40m are right-handed polarized. We only 
search for bursts in the left-handed polarization, but record the data from both 
polarisations. At NS26m, we search for bursts in the total intensity, i.e. the 
summation of two polarisations. The data is processed in real time, and only data 
with potential candidates of radio transients were recorded. 
The false alert probability 
of our  detection threshold is $10^{-7}$ (cf. \SEC{sec:rtpip}).

To save telescope resources, piggyback observations were carried out, i.e. we 
recorded the data while the telescopes were performing other 
observations. FRBs roughly distribute uniformly over the sky \citep{pvjb14}, 
so such an observation mode does not in principle reduce the event rate significantly.  
In practice, however, a significant fraction of observing time,
both at NS26m and KM40m, were allocated for pulsar observations. 
Observing along the Galactic
plane may lead to a lower event rate \citep{pvjb14}. We aimed to piggyback all 
observations at all available frequencies. By the time of writing this paper, 
due to the complexity in scheduling, we only piggybacked the 1.4 GHz 
observation at NS26m and 2.5 GHz observations at KM40m. 

\begin{table*}
	\centering
	\begin{threeparttable}
\caption{The specifications for the Nanshan and the Kunming radio telescopes.}

	\label{tab:teles}
	\begin{tabular}{ccccccccc} 
		\hline\hline
		Name & $\Delta \Omega^\ast$ & BW & $f_{\rm central}$ & $f_{\rm ch}$& $\Delta 
		t$& $G$ & $T_{\rm sys}$\\
						& sq. deg.	& GHz          & MHz                        &  
						MHz       & $\mu$s                                & K/Jy & K\\
		\hline
		NS26m	& 0.22 & 320 & 1.54 & 1.0 & 65 &  0.1   & 23\\
		KM40m & 0.04 & 140 & 2.24 & 1.0 & 65 &  0.23 & 70\\
		\hline
	\end{tabular}
	$^\ast$ Field of view\\
\end{threeparttable}
	\end{table*}

\subsection{Digital backends}
\label{sec:hard}

We recorded the data using home-brewed digital backends. It consists
of a Fast-Fourier transform (FFT) spectrometer and recording computer.
We built the FFT spectrometer based on the popular \textsc{Roach2}
platform\footnote{Reconfigurable Open Architecture Computing
Hardware:\\ \url{http://casper.berkeley.edu/wiki/ROACH2}}, where a
Virtex-6 family field programmable gate array (FPGA) of \textsc{Xilinx}
performs digital signal processing and data packetizing.
The FPGA gets the digitized signal from one 8-bit-5-Gsps analog-to-digital
Converter (ADC) ADC1x5000-8 produced by \textsc{E2V}. In each ADC, there
are four sampling cores. We configure the chip to sample the two
polarisations at 2$\times$2 Gsps. We generate the ADC clock signal
using \textsc{Valon} 5009 frequency synthesizer. The FPGA is fed
with the same clock to form the sampling clock.

In the FPGA, 1024 channels are created using a polyphase filter bank (PFB) and 
then correlated to compute
the coherency matrices, i.e. $\{XX^*, YY^*, {\rm Re}(XY^{*}), {\rm 
Im}(XY^{*})\}$, where $X$ and $Y$ are the complex voltage of the two 
polarizations, and $X^*$ and $Y^*$ are the corresponding
complex conjugates.  We integrate the coherency matrices for each
of 64 samples, which results in a spectrum with time resolution of
65 $\mu$s. The data is then packetized and transferred to the recording computer 
node with the user datagram protocol (UDP) over a 10-Gigabit Ethernet (10 GbE) 
optical link. \FIG{fig:FX_correlator} describes the hardware flowchart of our 
digital backend. 


\subsection{Realtime transient search and data analysis pipelines}
\label{sec:rtpip}

For piggyback observations, we do not want to store all the data,
since most of the data will not contain FRB signals. Also, in order to keep the 
data volume manageable, the data recording
rate is limited and only data of candidates are
stored. To do so, we implemented a realtime searching and data analysing 
pipeline. 

Our realtime transient search system is called `Burst Emission Automatic Roger' 
(\BEA), which has three major components, 1) the data processing manager (DPM), 2) 
the data buffering component (DBC), and 3) the data analysis component (DAC). The 
DBC captures the UDP packets from the 10 GbE optical link and buffers the data 
in the shared memory. The DAC firstly mitigates RFI, then de-disperses the 
signal at a given set of DM grids, searches for pulses using matched filter in 
the de-dispersed time series, and clusters the candidates to identify the burst 
events.  There are multiple copies of the DAC processes running in the computer node 
to parallelize the data processing. The DPM allocates the shared memory on the data 
recording machine, and creates globally visible flags for each DAC task to 
coordinate the work.
We will explain each parts separately in the following sections.

\subsubsection{DBC, Data buffering}
\label{sec:databuf}
When the \BEA\ system starts, the DPM allocate 8 segments of shared memory in 
the data recording machine. DPM also creates globally visible flags for each of 
the 8 segments. Each flag has five states, namely, `empty', `writing', `ready', 
`reading', and 'saving'. After the shared memory is created, the initial flags 
are all set to `empty' by the DPM. 

The DBC captures the data from the 10 GbE optical link. It searches for the
flags of `empty'. Once an `empty' segment of the shared memory is found,
the DBC changes the flag to `writing' and starts to store the captured
data into the corresponding memory block. The DBC changes the flag to `ready' and 
seek for the next empty memory block, when the memory block is filled fully. The 
`ready' flag notifies the DAC, then the DAC starts the data analysis after 
modifying the flag to `reading'.
This prevents the DBC or other instances of DAC to interfere with the
data processing. After DAC processes  the data, the two possible flags are
`saving' or `empty'. If DAC finds candidates in the buffered data, the
`saving' flag is assigned, and DPM is notified to save the data to the hard 
disk. If no candidate is detected, the `empty' flag is
given, and memory block will be re-written by the DBC. Using this flag scheme, 
as far as the candidates rate is limited, we can
even save the baseband data with a limited storage resource. For
example, if we pick up candidates at a 5\% probability, 500 Mbps
file saving speed can handle incoming data with the rate of 10 Gbps. It
is thus possible to do baseband data recording with only one computer
node for FRB searching.

\subsubsection{DAC, RFI mitigation}

The first task of DAC is to mitigate the RFI signals in the data. Beside the common 
practice of zapping the channels with known persistent RFI \citep{FB01}, we 
also perform the RFI mitigation called \emph{zero-DM matched filter}, which effectively removes RFI of terrestrial origins (signal with nearly zero DM). 

The zero-DM matched filter (ZDMF) is an improved version of the zero-DM filter (ZDF) developed by \citet{Eatough2009}. The ZDF subtracts the zero-DM time series from the data, which significantly reduces the local RFI. In fact, the initial application of ZDF helped in the discovery of four new pulsars \citep{Eatough2009}, and has later been applied in several pulsar or single-pulse surveys \citep{Eatough2013mn, Keane2010mn, Rane2016mn, Patelapj}. The ZDMF estimates the waveform of the zero DM signal similarly to the ZDF, but subtracts only the corresponding contribution from each channel. Such a modification reduces the over-subtraction, when dealing with narrow-band RFI.

Similar to the zero-DM filter, the zero-DM time series (i.e. the `audio' signal) 
is estimated by de-dispersing the original data at zero DM value. This is done 
by adding data of all frequency channels. The zero-DM waveform is denoted here as 
$\VEC{s}_{\rm dm=0}$.  We then estimate the corresponding contribution from each 
channel, i.e.  we find the baseline $\beta_i$ and the scale factor $\alpha_i$ 
for each channel such that the residual of fitting the zero-DM waveform to the 
given channel is minimized.  The residual $\chi^2$ is defined as 
\begin{equation}
	\chi^2=(\VEC{s}_{i}-\alpha_i \VEC{s}_{\rm dm=0}-\beta_i)^2\,,\label{eq:dmres}
\end{equation}
where $\VEC{s}_i$ is the time series of the $i$-th channel. The factor 
$\alpha_i$ can be found analytically as
\begin{equation}
	\alpha_i = \frac{\VEC{s}_{\rm dm=0} \cdot \VEC{s}_i  -\frac{1}{N}\sum s_i \sum 
	s_{\rm dm=0}}{\VEC{s}_{\rm dm=0} \cdot \VEC{s}_{\rm dm=0} -\frac{1}{N}\sum 
	s_{\rm dm=0} \sum s_{\rm dm=0}}\,.
	\label{eq:alpha}
\end{equation}
Here, $\cdot$ is the inner product of the one dimensional time series. The symbol 
$\sum$ indicates the time domain summation with a total of $N$ data points.
With the $\alpha_i$, we remove the RFI from the data using
\begin{equation}
	\VEC{s}'_i=\VEC{s}_i-\alpha_i \VEC{s}_{\rm dm=0}\,,
	\label{eq:rfimit}
\end{equation}
where the new time series $\VEC{s}'_i$ is the data of the $i$-th channel with 
RFI removed. The DC-offset $\beta_i$ is not removed here, because it doesn't affect 
the pulse detection.

\begin{figure}
	\includegraphics[width=\columnwidth]{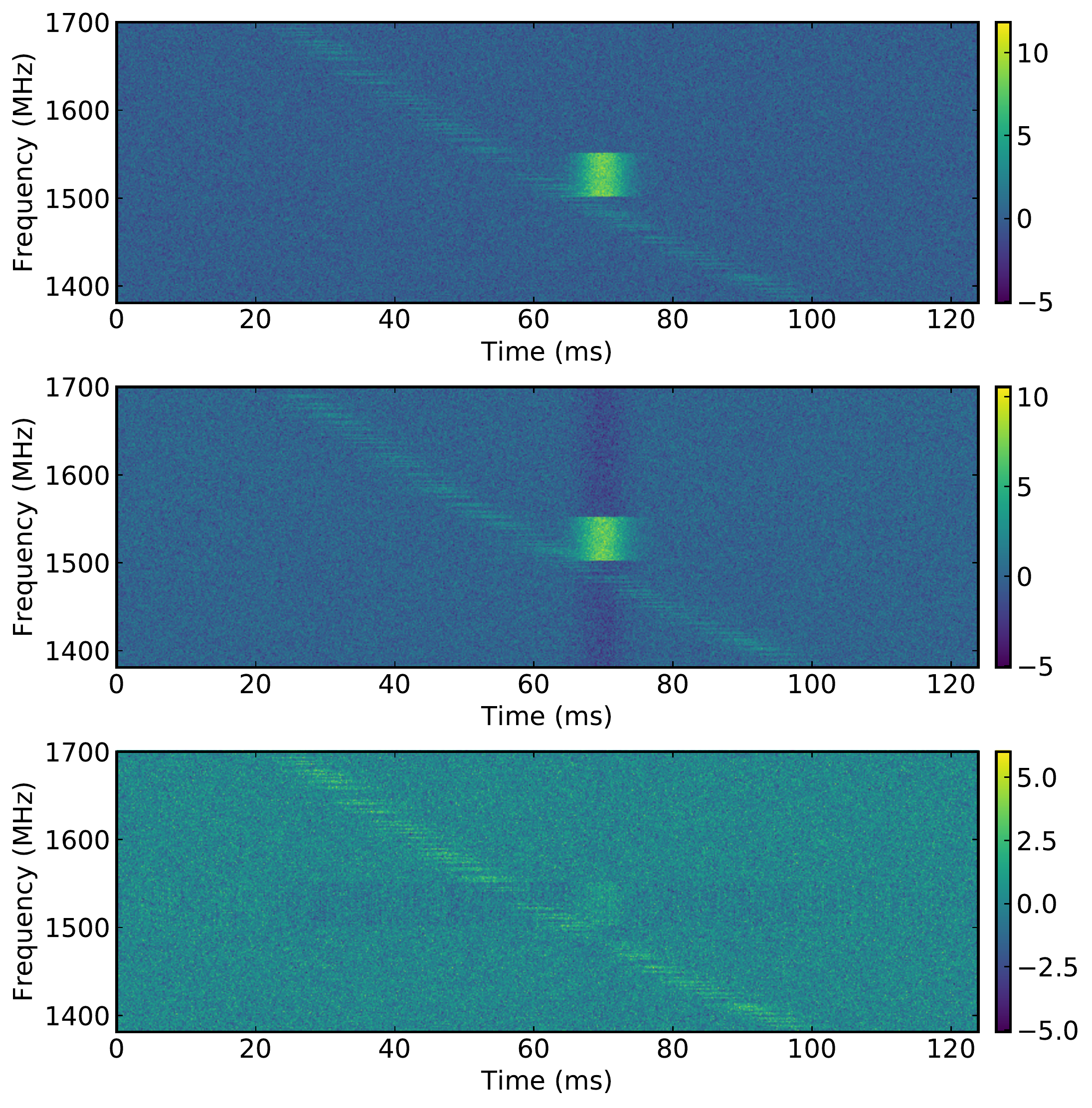}
		\caption{Examples for RFI mitigation algorithms. The top panel shows a 
		simulated data set containing a highly frequency-modulated signal and a narrow-band RFI extended over a short 
		duration, where the x-axis is the time, 
		the y-axis is the observing frequency, and the 
		colour/gray scale indicates signal strength. The middle panel shows the result after
		RFI mitigation using the `zero-DM' technique \citep{Eatough2009}, where dips 
		of intensity is introduced for channels containing no RFI. The bottom panel 
		is the result using our `zero-DM matched filter'. The figures show that 
		the current RFI mitigation works rather well, and only low level artifacts 
		are left in the channels affected by the RFI.}
    \label{fig:rfi_mitigation}
\end{figure}

Examples of our RFI mitigation are shown in \FIG{fig:rfi_mitigation}. The 
ZDMF can effectively remove RFI. The figure also shows the comparation between
the ZDMF and ZDF.  

\subsubsection{DAC, De-dispersion and burst searching using matched filter}

After RFI mitigation, we search for bursts in the data using the matched 
filter.
Similarly to any matched filter for signal detection, if the parameter of the 
matched filter is slightly away from the `true' value, 
there will be loss of signal-to-noise ratio (SNR).  
That is, if the true parameters of the FRB (DM, burst epoch, pulse width) is off 
the matched filter parameter grid, the detection probability becomes lower.  As 
we will show below, we designed a nonlinear parameter searching grid, 
such that the SNR loss of \BEA\ is always smaller than a preset threshold. In 
this way, we minimize the number of trials needed for the matched filter.

\begin{figure}
	\includegraphics[width=3.2in]{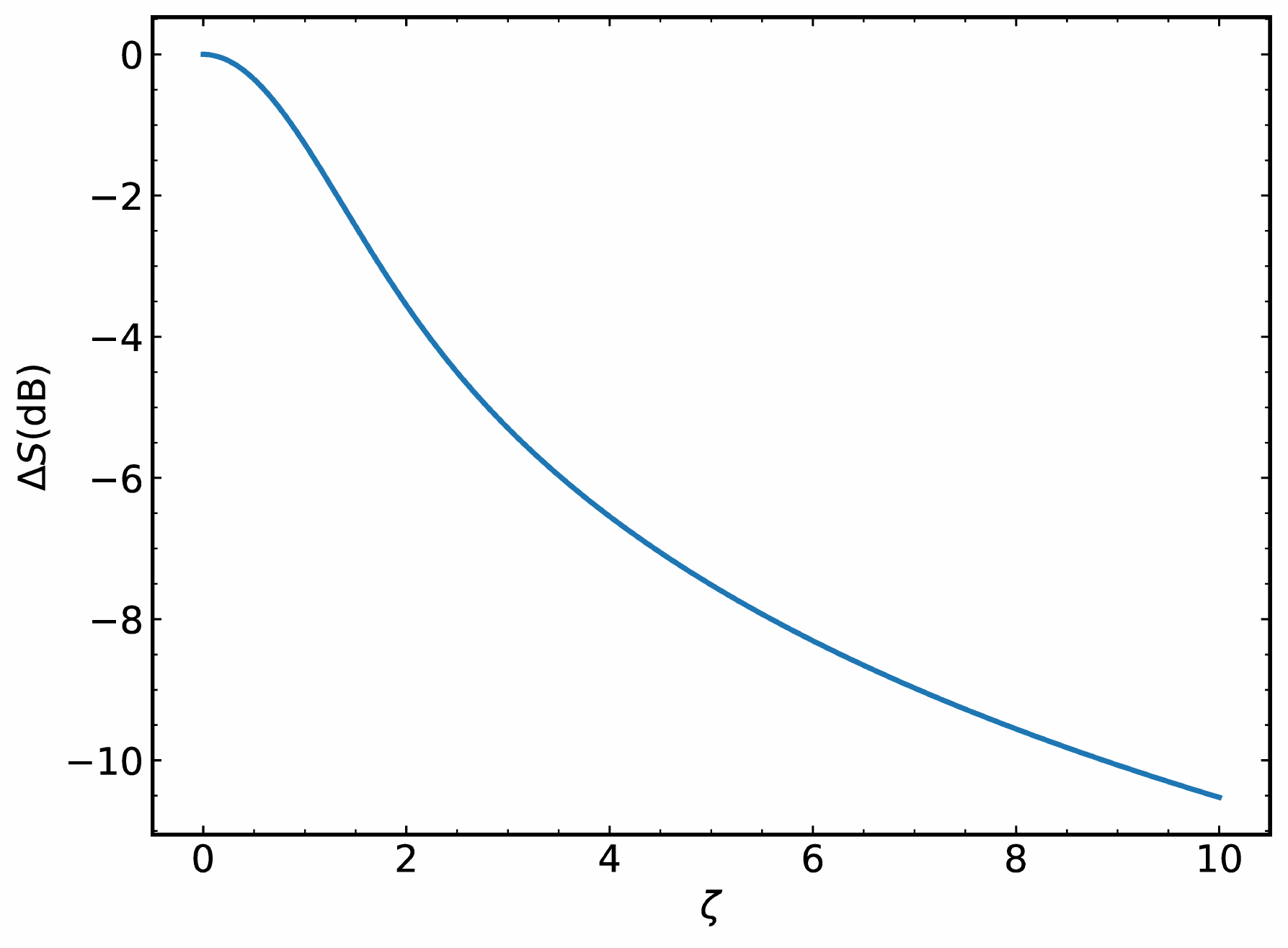}
	\includegraphics[width=3.2in]{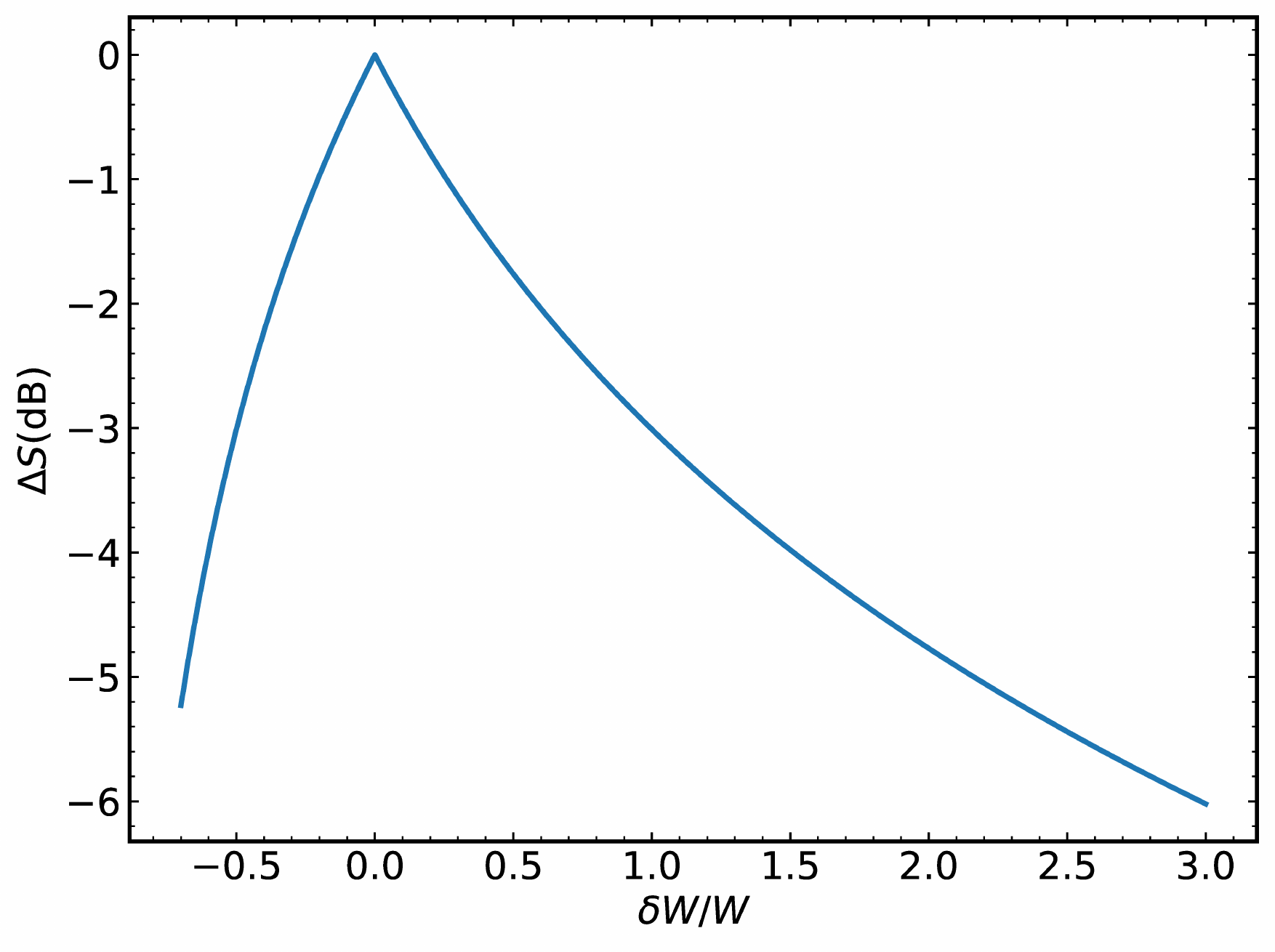}
		\caption{The SNR loss as the function of parameter mismatching. The y-axis 
		is the SNR loss using a unit of log-10 based decibel. The upper panel is for 
		the SNR loss due to mismatch of true and searching DM.  The x-axis is the 
		parameter $\eta$ defined in \EQ{eq:zeta}. The lower panel is the SNR loss 
		as function of width mismatch, where x-axis $\delta W/W$ is the fraction 
		width of the pulse width mismatch.}
		\label{fig:snrlossmis}
\end{figure}

In order to search for radio bursts of celestial origins,
we need to align the data by correcting for the dispersion effect
(c.f. \EQ{eq:dispersion}). If the DM trial value is off the true
value, the recovered burst becomes wider, extra noise is added to
the burst, and the SNR of the burst signal reduces. If the DM offset
is $\delta {\rm  DM}$, the ratio between detection SNR and expected SNR
is  \citep{Cordes2003} \begin{equation}
	\frac{\rm SNR}{\rm
	SNR_0}=\frac{\sqrt{\upi}}{2}\zeta^{-1}\mathrm{erf}\zeta\,,
	\label{eq:dmsnrreduc}
\end{equation} where $\rm SNR_0$ will be the SNR, if the true DM is used
in searching and SNR is the observed SNR. The $\rm erf$ is the error
function, and $\zeta$ is the ratio between the time delay caused by the
DM offset and pulse width, i.e.  $\zeta$ is \begin{equation}
		\zeta =
		6.91\times10^{-3}\delta\mathrm{DM}\frac{\Delta\nu_\mathrm{MHz}}{W_\mathrm{ms}\nu_\mathrm{GHz}^3}\,,
	\label{eq:zeta}
\end{equation} where $W_{\rm ms}$ is the pulse width in units of
millisecond, $\nu$ is the observing central frequency in GHz, and $\Delta
\nu_{\rm MHz}$ is the bandwidth in MHz.
If one uses decibel-scale SNR, denoted as $\cal S$,
defined as ${\cal S} \equiv 10 \log_{\rm 10} {\rm SNR}$ ,
\EQ{eq:dmsnrreduc} which describes the SNR loss can be well
approximated by
\begin{equation}
	\Delta {\cal S}_{\rm dB}\simeq -\frac{5}{3} \zeta^2\,.
	\label{eq:siglossdmdb}
\end{equation} For our 1.5-GHz observation of 300\,MHz bandwidth,
the maximum allowable 15\% SNR loss, i.e. -0.7 dB, constrains the
DM searching grid to be uniform with a step of $1\ \rm cm^{-3}\,
pc$. For high DM, the SNR loss caused by the intrachannel smearing will be 
dominant over the DM trial errors. One can use larger DM steps to reduce 
the computational cost. However, in order to also study the
interferences with dispersive signatures and to simplify the data reduction, our 
DM trial grids span from 200 to 3000 $\mathrm{cm^{-3}\,
pc}$ with 1 $\mathrm{cm^{-3}\ pc}$ increments. The SNR loss of the DM
mismatch is plotted in \FIG{fig:snrlossmis}.

As indicated by \EQ{eq:siglossdmdb}, a preset SNR loss requires an uniform grid 
for DM trials. We use the subband de-dispersion algorithm \citep{Magro2011}
to de-disperse the data. The algorithm first divides the total $N$ channels 
into $N_{\rm sub}$ subbands, and de-disperses each subband over a coarse DM 
grid. Then the de-dispersed subband data is combined by another layer of 
de-dispersion to form the final de-dispersed 1-D time series on finer grid.  For 
the optimal choice, this algorithm roughly speeds up the computations by factor 
of $\sqrt{N_{\rm sub}}$.

The in-channel dispersion smearing also introduces SNR loss. Such an effect is 
very similar to the DM mismatching loss. If we substitute the bandwidth $\Delta 
\nu$ with channel width and $\delta \rm DM$ with the DM in \EQ{eq:zeta}, we can 
compute the SNR loss due to channel smearing. Our backend records with 1 MHz 
channel width, this leads to a maximum $-0.7\, {\rm dB}$ loss for pulse signal 
of 5 ms with $\rm DM=2000$.

We then search for burst signals in the de-dispersed 1-D time series using the 
matched filter technique. We \emph{assume} the burst can be approximated by a 
square shaped wave, i.e. the template for the filter is
\begin{equation}
	h(t; t_0,w)=\left\{\begin{array}{l}
		A, \,{\rm if\,} |t-t_0|\le W\\
		0, \,{\rm otherwise}
	\end{array}\right.,\label{eq:sqwtem}
\end{equation}
where $A$, $t_0$, and $W$ are the amplitude, centre epoch and width of the 
square-shaped burst.  The `most powerful test', a statistical tool to detect 
signal with maximum detection probability with fixed false alarm probability,  
comes from the likelihood ratio test \citep{Fisz63}. The detection statistic 
$S$ is the logarithmic likelihood ratio between the cases of having and not 
having the signal. As shown in the \APP{sec:appliktst}, for the Gaussian noise case, 
one has \begin{equation}
	S=\frac{\VEC{s}^2-(\VEC{s}-\VEC{h})^2}{2\sigma^2}\,
	\label{eq:dets}
\end{equation}
where $\sigma$ is the standard deviation of the noise in 1-D time series. With 
the square wave filter (\EQ{eq:sqwtem}), \EQ{eq:dets} is reduced to 
\begin{equation}
	S=\frac{1}{N_{\rm box} \sigma^2} \left(\sum_{|t-t0|\le W} s(t)\right)^2\,,
	\label{eq:1ddets}
\end{equation}
where $N_{\rm box}$ is the number of data points in the time span where 
$|t-t_0|\le W$. The likelihood ratio statistic $S$ is nothing but the square of 
the burst signal SNR, i.e.
\begin{equation}
	{\rm SNR}=\sqrt{S}\,.
	\label{eq:snrs}
\end{equation}

We set the detection threshold $\gamma_0$, such that we will only record data 
when $S\ge \gamma_0$. For the null hypothesis, i.e. there is no burst in the 
data, the distribution of $S$ follows a $\chi^2$ distribution with one degree of 
freedom. The corresponding false alarm probability $P_{\rm FA}$ of the 
given threshold $\gamma_0$ is thus
\begin{equation}
	P_{\rm 
	FA}=\rm{erfc}\left(\sqrt{\frac{\gamma_0}{2}}\right)\simeq\sqrt{\frac{2}{\upi 
	\gamma_0}}{\rm e}^{-\frac{\gamma_0}{2}}\,,
	\label{eq:falseprob}
\end{equation}
where function erfc is the complementary error function defined by ${\rm 
erfc}(x)\equiv1-{\rm erf}(x)$. The approximation in the equation above is valid when 
$\gamma_0\gg 2$. In our searching pipeline, we use threshold of $P_{\rm FA}\le 
10^{-7}$, i.e. $\gamma_0\ge28$. The approximation is therefore good enough to calculate the false alarm probability. 

When there is a burst in the signal with a given SNR, the detection probability 
corresponding to the detection statistic $S$, i.e. the probability to get a $S$ 
larger than the threshold $\gamma_0$ is
\begin{equation}
	P_{\rm D}=	1-\frac{1}{2} \left[{\rm erf} \left( \frac{{\rm 
	SNR}+\sqrt{\gamma_0}}{\sqrt{2}}
	\right)-{\rm erf} \left( \frac{{\rm 
	SNR}-\sqrt{\gamma_0}}{\sqrt{2}}
	\right)\right]\,.\label{eq:detprob}
\end{equation}

\begin{figure}
 \includegraphics[width=3.2 in]{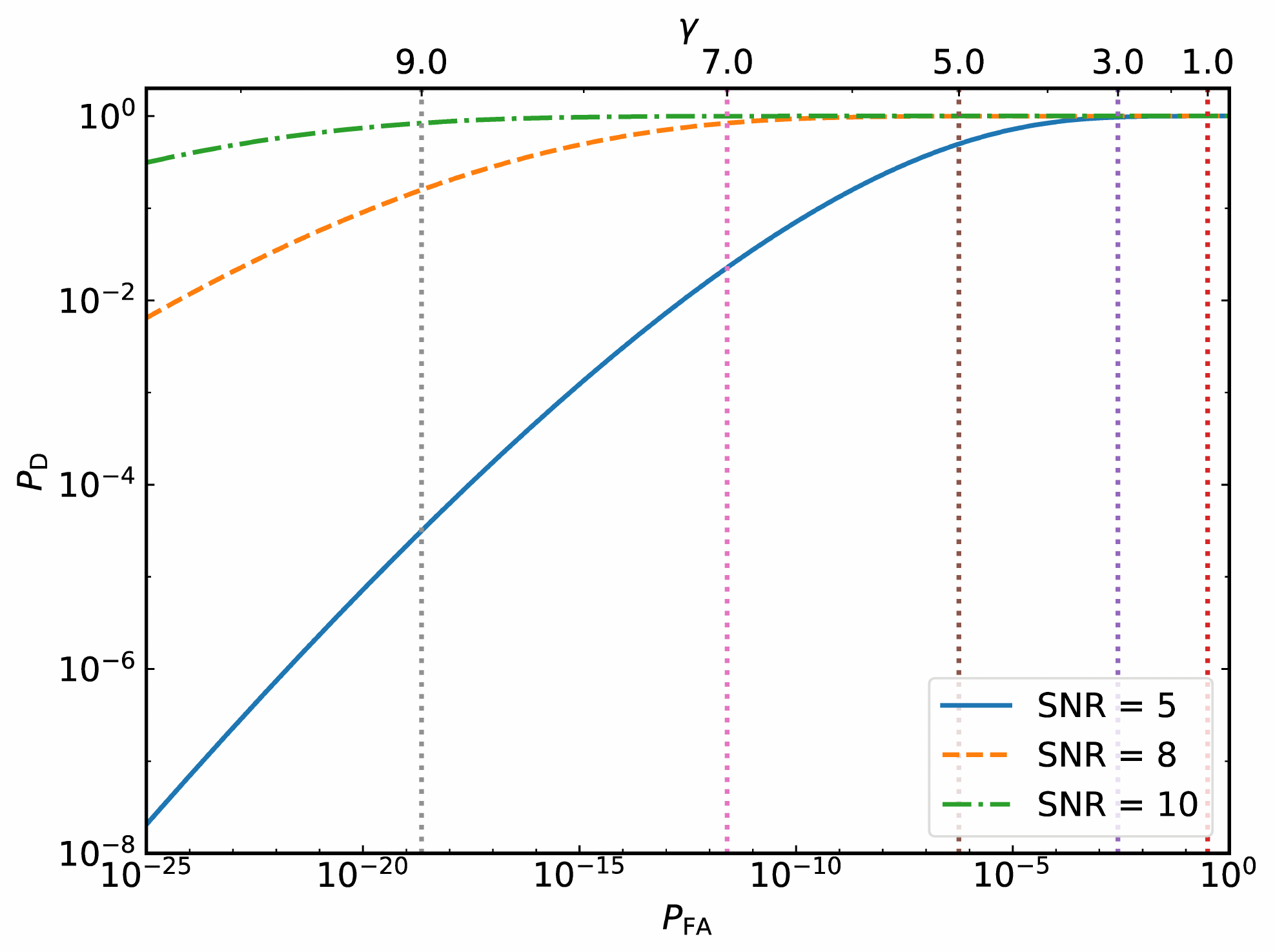} \caption{ROC curves for the matched filter 
 using a square pulse waveform. The x-axis is the false alarm probability, and 
 y-axis is the detection probability. The SNR of each curve is labelled in the 
 figure. On the top of the panel, we also labelled the corresponding statistical 
 threshold $\gamma_0$.}
\label{fig:ROC} \end{figure}

With the false alarm probability and detection probability, the statistical 
performance of the matched filter can be evaluated using the `receiver operating 
characteristic' curves (ROC curves), it is the relation between $P_{\rm D}$ and 
$P_{\rm FA}$ parameterized using threshold $\gamma_0$.  For reader's reference, 
the ROC curves of the matched filter described here is given in \FIG{fig:ROC}.  

With the matched filter, we can now turn to problem of the equal-SNR-loss grid 
design for pulse width parameters $W$. The SNR loss with slightly wrong $t_0$ 
and $W$ becomes
\begin{equation}
	\frac{\rm SNR}{\rm SNR_0}=\left \{\begin{array}{l}
		\frac{W}{W+\delta W} , {\, \rm for\,} \delta W\ge0
\,, \\
		\frac{W+\delta W}{W}, {\, \rm for\,} \delta W<0\,.
	\end{array}\right.
	\label{eq:siglossw}
\end{equation}
where $\delta W$ is the mismatch of the pulse width. 
By setting the maximum SNR loss as 
15\%, the allowable grid for $W$ spans from $0.85 W$ to $1.17 W$. In this way, 
the equal-SNR-loss for the $k$-th grid should be a geometric series of $W_k=1.37^k 
W_{\rm min}$, and $W_{\rm min}$ is the minimum pulse width in the searching.  
Our minimum searching grid is 0.5 ms, and using only 12 grid steps cover the 
width search from 0.5 ms to 20 ms with a maximum SNR loss of 15\%. The SNR loss 
and function of pulsar width mismatch is plotted in \FIG{fig:snrlossmis}.

The equal-SNR-loss grid for pulse epoch $t_0$ is a function of pulse width. If 
the equal-SNR-loss grid would be used for $t_0$, we would need to adjust it according 
to the pulse width grid. Luckily, due to a very efficient  method to compute the 
statistic $S$, we can use a uniform-grid steps of 0.5 ms for $t_0$ in our 
searching. The $S$ is computed using the running averaging of data, which can be 
done very efficiently by subtracting one earlier data point and adding one new data 
point. In this way, the complexity of applying the square wave matched filter 
becomes of order $O(n)$ time complexity rather than well-known time complexity 
of order $O(n\log{}n)$ for applying the filters using the fast Fourier transform.

\subsubsection{DAC, Clustering of candidates}

By using de-dispersion, and matched filter, \BEA\ computes the detection 
statistic $S$ as a cube on a 3-D parameter grid spanned by DM, $W$ and $t_0$.  
\BEA\ reports detection, if the statistic $S$ is larger than the threshold 
$\gamma_0$.  However, we note that naively reporting all the candidates with 
$S\ge \gamma_0$ is rather inefficient, as we need to remove the duplicated 
candidates.

Indeed, if the signal is strong, even in trials were
with parameters mismatching the central 
values, the computed statistic still report detection. There will be many 
candidates clustering around the central peak of $S$. All those candidates in 
such a cluster are basically the same burst signal, but `found' at slightly 
different parameters. We use the method called \emph{candidate clustering} to 
remove such a redundancy.

Our recipe of candidates clustering is similar to the cleaning algorithm in the 
radio interferometry. The steps are as follows: 1) Find the grid with the
highest value of $S$ in the detection statistics cube.  2) Find the neighbours of 
the grid with highest $S$. Here the `neighbours' are the grids contacting the 
given central grids.  3) Find the neighbours of the neighbours, where each outer 
layer of neighbours has lower value of $S$ compared to the inner layer of 
neighbours. 4) Find the boundaries of all the neighbourhood region, where the 
boundaries are either confined by $S\ge \gamma_0$ or the requirements of 
monotonically decreasing $S$, 5) Report the central $S$ as one candidate,
removing such the connected grid region, and repeating the procedure from the 
step 1.  Here the monotonically decreasing $S$ ensures that interesting 
candidates do not get shadowed by other bright candidates nearby in the 
parameter grids.

With this procedure, we cluster the related candidates and the duplication of
candidates is suppressed. Examples of the results will be given in the next 
section.

\subsubsection{Test of pipelines}

Our pipeline, from the home-brewed digital backends to the realtime searching is 
tested both in laboratory and on telescope site. 

\begin{figure*}
\includegraphics[width=\columnwidth]{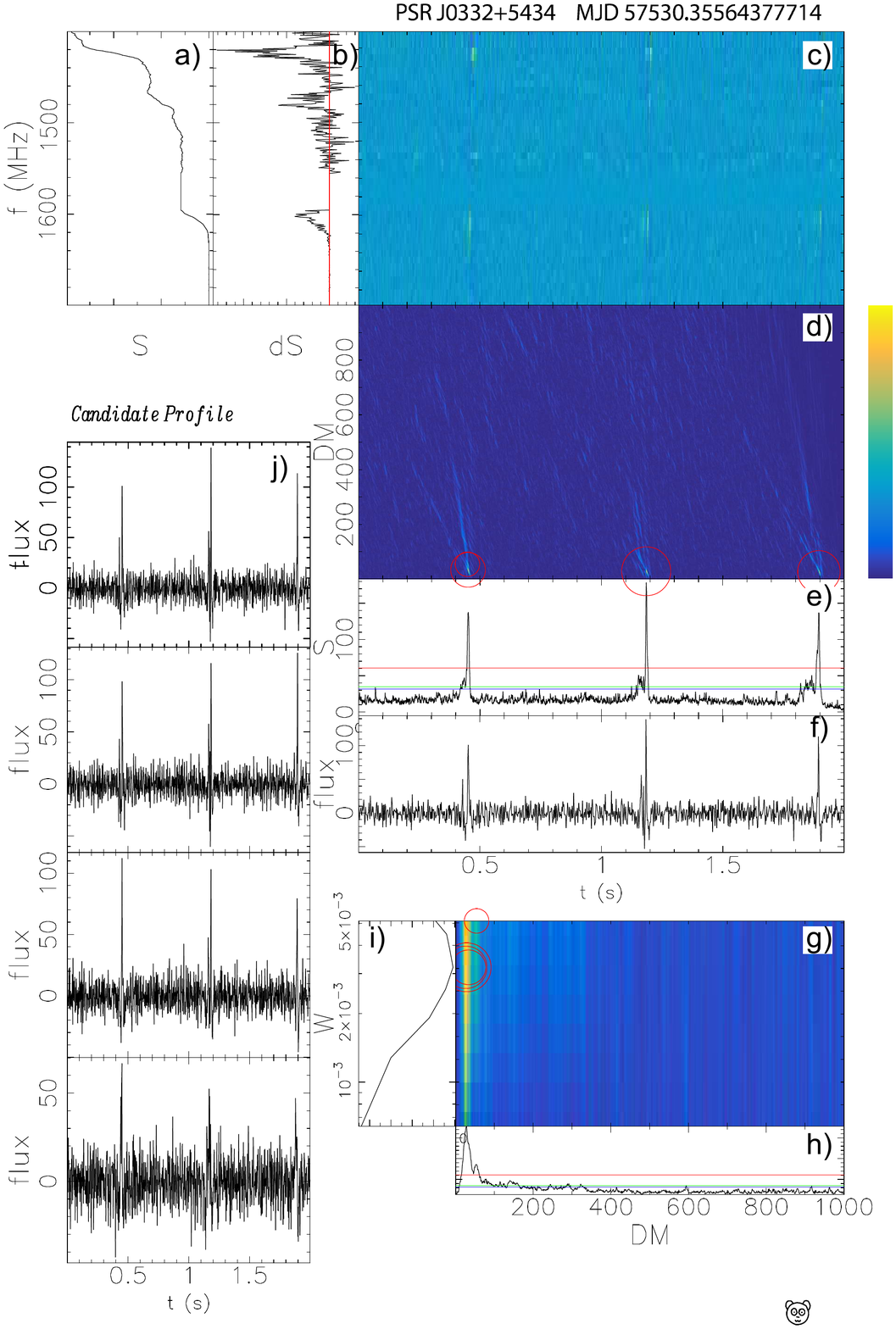}
\includegraphics[width=\columnwidth]{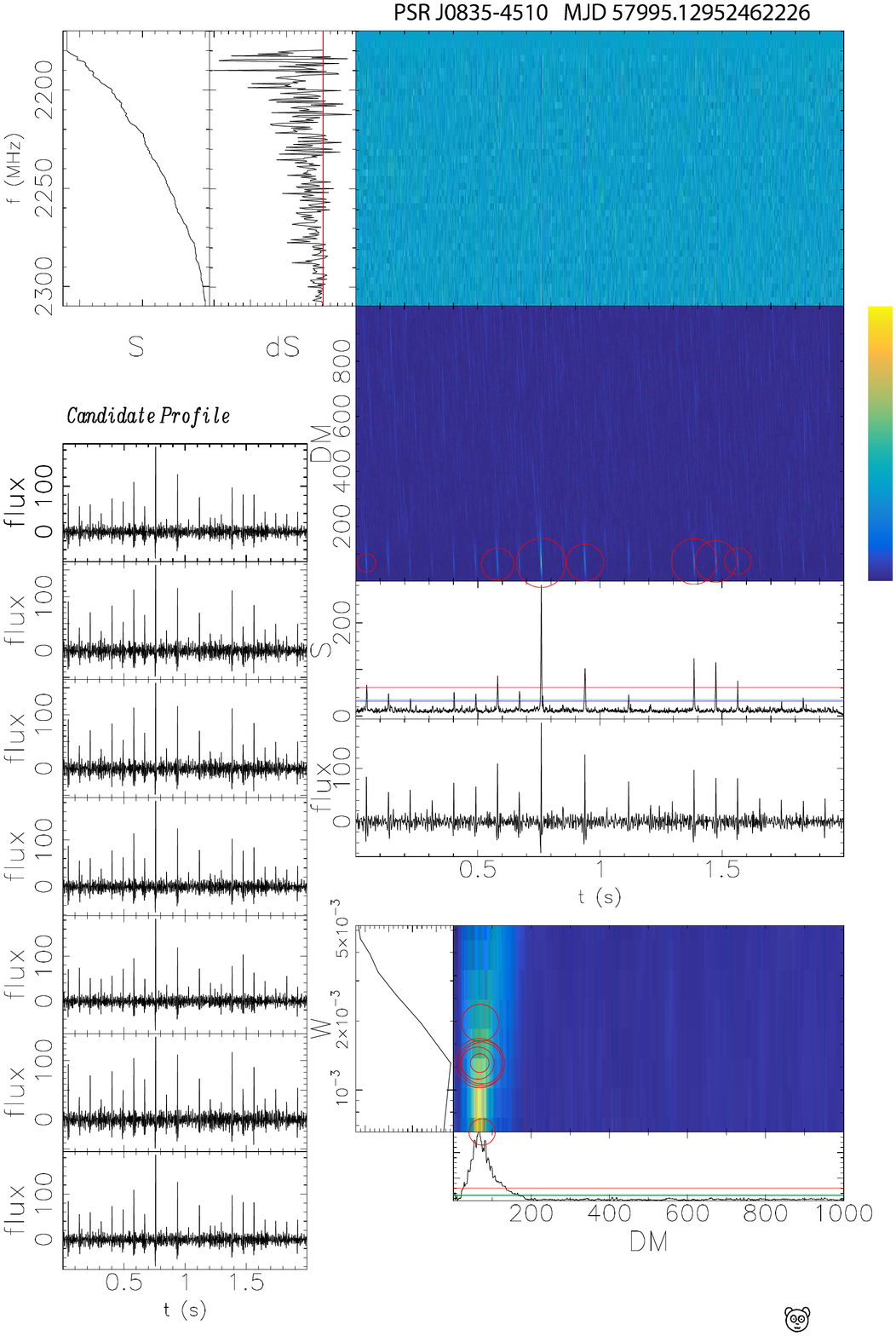}
\caption{Left panel: The single pulse of PSR J0332+5434 detected at the 
NS26m. Right panel: The single pulse of J0835$-$4510 (Vela) detected at the 
KM40m. In the left panel, we also label each subplot with alphabet letters (a to h). 
The meaning of each subplots are, a) the integration of $S$ as function of 
frequency,  b) the contribution of $S$ from each channel, c) filterbank data, 
the intensity as a function of frequency (y-axis) and time (x-axis), d) The $S$ 
as a function of DM (y-axis) and time (x-axis), e) $S$ as function of time, where 
red, green and blue horizontal lines (online version) correspondents to the 
$P_{\rm FA}$ of $10^{-7}$, $5\times 10^{-3}$, and $0.3$ respectively, f) 
de-dispersed 1-D time series of the highest $S$, g) $S$ as a function of width 
$W$ (y-axis) and DM (x-axis),  h) $S$ as function of DM, i) $S$ as function of 
$W$, j) the pulse profile of each individual candidate. In the subplot c) the 
red circle indicates the location of each local maximum of $S$, which is the 
candidate reported in i).  The size of the red circles are specified by the 
corresponding value of $S$, i.e.  the SNR. As one can see, besides the local 
maxima indicated by the red circles, there are grids with values higher than 
the threshold. As shown in the subplot d) the $S$ around peak could still be higher 
than the red line. Our clustering algorithm combined the region around the peak
and reduced the number of candidates reported in the subplot i).}
\label{fig:pipeline_test}
\end{figure*}

In the lab, we simulated the FRB radio frequency signal by modulating a wideband 
noise signal with low frequency pulses. The signal is fed to the 
backend and we check if the \BEA\ correctly detects the injected pulse. 
At the telescope sites, we tried to catch the single pulse stream 
from bright pulsars. We observed PSR B0329+54 at NS26m and the Vela pulsar at 
KM40m for 5 minutes. During the time, bright single bursts were detected and 
recorded. The candidate plots generated by the realtime searching pipeline is 
shown in \FIG{fig:pipeline_test}. The candidate plot also contains extra 
information to aid the users to do further inspection.  The meaning of each 
panel is explained in the figure caption. 

\section{Expectation of searching sensitivity and event rates for FRBs}
\label{sec:eventrate}

In this section, we estimate the expected event rate of our experiment. At the 
time the current FRB searching project started, the event rate estimation for 
NS26m and KM40m was very uncertain. Most of the FRBs at that time were 
detected by the Parkes telescope. The sensitivity of Parkes is higher than both 
the NS26m and KM40m, but there is lack of information for the close-by FRB 
population, i.e. FRBs with higher flux. Thanks to the ASKAP survey 
\citep{SMB18}, we can now do a better estimation for the FRB event rate.

The minimum detection amplitude of FRB events is
\begin{equation}
	S_\mathrm{min}=\beta \frac{\sqrt{\gamma_0}\ T_\mathrm{sys}}{G 
	\sqrt{\mathrm{BW} \tau N_\mathrm{p}}},
	\label{eq:sensitivity}
\end{equation} where $S_\mathrm{min}$ is the minimum detectable flux for a given 
statistical threshold $\gamma_0$, $\beta\simeq 1$ is the digitisation factor, 
$\mathrm{BW}$ is the bandwidth, $N_\mathrm{p}$ is the number of polarisations, 
$\tau$ is the pulse width, $T_\mathrm{sys}$ is the system temperature
and $G$ is the telescope gain. 
By choosing 3 ms as the reference width of FRB, the flux 
thresholds for SNR > 7 for NS26m and KM40m are 1.1 Jy and 3.3 Jy, respectively, using telescope parameters in \TAB{tab:teles}.

\begin{figure}
	\includegraphics[width=\columnwidth]{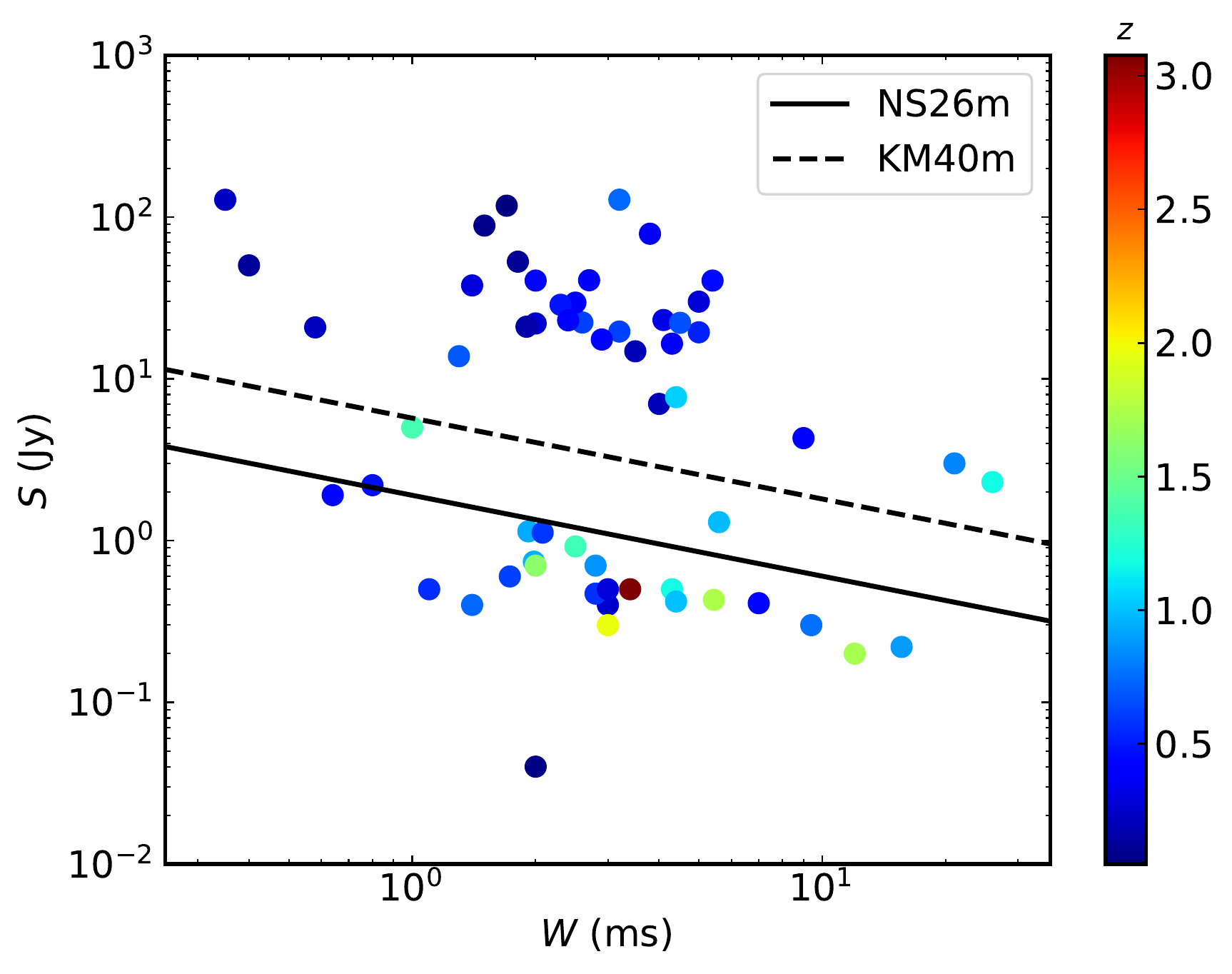}
		\caption{The currently known FRBs (coloured dots) together
		with the 7-$\sigma$ sensitivity of NS26m (solid line)
		and KM40m (dashed line). The x-axis is the FRB pulse width
		and the y-axis indicates the observed flux. The colour of
		each dot indicates the redshift estimated using 
		the maximum-likelihood estimator of \citet{lll18}. The data of known FRBs
		comes from the FRBCAT \citep{Petroff2016}.  More than
		half of the known FRBs would be detectable with both
		the telescopes we used.}
    \label{fig:sensitivity_width}
\end{figure}

\begin{figure}
	\includegraphics[width=\columnwidth]{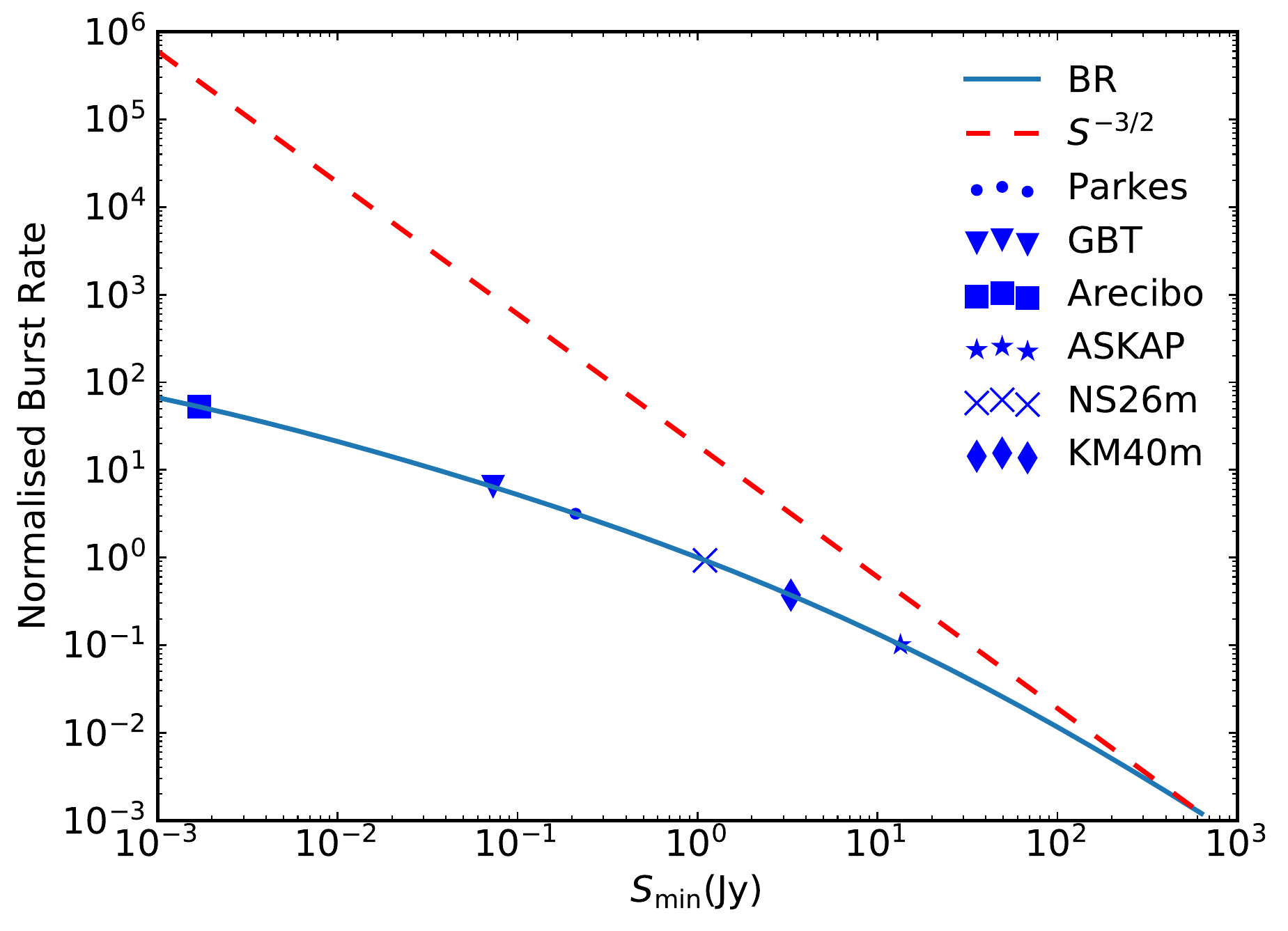}
	\caption{The normalized burst rate (see \EQ{eq:eventrate}) as function of 
	the minimum detectable amplitude. The solid blue curve is the computation made in 
	this paper. Here, we normalize the burst rate arbitrarily to $S_{\rm min}=1$ 
	Jy, as the event rate density of FRB in the Universe is rather uncertain. The 
	red dashed curve is for $S_{\rm min } ^{-3/2}$, i.e. the expected normalized 
	burst rate in flat Euclidean space. One expects that the Euclidean 
	approximation will be valid for FRBs located near the Earth. Indeed, the slope 
	of the solid blue curve approaches $-3/2$, when $S_{\rm min}$ becomes 
	large and observers can only see the near-by FRBs. We also compute the 
	corresponding value of burst rate for a few other telescopes, which have 
	detected FRBs, as indicated by the symbols on the solid blue curve. Telescope 
	parameters are from \citet{lll18}. Note that we use a beam-central gain of 10 
	${\rm K/Jy}$ for Arecibo in the current computation to account for the 
	sidelobe effects.}
		\label{fig:BRS}
\end{figure}

As shown in \FIG{fig:sensitivity_width}, more than half of the known FRBs would 
be detected with NS26m and KM40m. Most of these detectable FRBs have redshift 
$z\le 1$. Integrating over the cosmological comoving volume, the \emph{expected} 
full-sky burst rate ($\rm BR$) in units of 1 per day per $4\pi$ solid angle is
\begin{equation}
	{\rm BR}=\int_{0}^{\infty} \phi(L) dL \int_0^{z_{\rm max}(L)} 
	\frac{1}{1+z}\frac{dV}{dz} dz\,,
	\label{eq:eventrate}
\end{equation}
where $\phi(L)$ is the event rate luminosity function of FRBs in units of 1 per 
day per Mpc$^3$ per luminosity ($L$) in the comoving frame. $z$ is the cosmological 
redshift. The $1+z$ factor in the denominator comes from the reduction of event 
rate for the observer on Earth due to the cosmological time dilation. The 
differential comoving volume ($dV/dz$) is
\begin{equation}
	\frac{dV}{dz}=\frac{c}{H_0} \frac{1}{E(z)} \left(\frac{c}{H_0}\int_{0}^{z} \frac{1}{E(z)} dz 
	\right)^2\,,
	\label{eq:dvdz}
\end{equation}
where the Hubble constant $H_0 = 67.8\,{\rm km s^{-1} Mpc^{-1}}$ 
\citep{Planck18}, $c$ is the light speed, and function $E(z)$ is the logarithmic 
time derivative of cosmic scale factor.  With matter density $\Omega_{\rm 
m}=0.308$ and cosmological constant $\Omega_{\Lambda}=0.692$ \citep{Planck18}, 
one has
\begin{equation}
	E(z)=\sqrt{\Omega_{\rm m}(1+z)^3+\Omega_{\rm Lambda}}\,.
	\label{eq:cosE}
\end{equation}
The function $z_{\rm max}(L)$ is the maximum redshift of detectable FRBs with 
intrinsic luminosity $L$, i.e. it is defined implicitly via \begin{equation}
	L=4\pi r_{\rm L}(z_{\rm max})^2 {\rm BW} S_{\rm min}\,,
	\label{eq:zmax}
\end{equation}
where BW is the bandwidth of receiver at the Earth. $r_{\rm L}$ is the 
luminosity distance defined as \begin{equation}
	r_{\rm L}(z)=\frac{c(1+z)}{H_{\rm 0}}\int _0^{z}\frac{1}{E(z)} dz\,.
	\label{eq:rl}
\end{equation}
The normalized burst rate as function of the minimum detectable amplitude is shown in \FIG{fig:BRS}.

Since the number density of FRB per comoving volume is not known
\citep{lll18}, we can not directly compute the expected event rate ($\rm BR$) 
for a given telescope.  We have to use the observed event rate at Parkes and
the ratio between the expected event rate at our two telescopes
to estimate the event rate at
KM40m or NS26m. Here the ratio between the expected event rates of the two
telescopes does not depend on the FRB number density anymore, as the
number density is a normalisation factor and cancels out in the ratio.
Denoting the observed event rate at the two telescopes as $\rho_{1}$ and $\rho_2$ 
(counts per day per full sky), we have
\begin{equation}
	\rho_2=\frac{\rm BR_2}{\rm BR_1}\rho_1\,,\label{eq:rho}
\end{equation}
Clearly, the ratio $BR_{\rm 2}/BR_{\rm 1}$ becomes independent of the FRB number 
density, i.e. independent of $\int_0^{\infty} \phi(L) dL$.

The detection rate ($dN/dt$, i.e. counts per day) of a given telescope is 
\begin{equation}
	\frac{dN}{dt}=\frac{\rho N_{\rm b } \Delta \Omega}{4\upi}\,, \label{eq:detrate}
\end{equation}
with $\Delta \Omega$ the solid angle of telescope main beam size, and $N_{\rm 
b}$ the number of beams. Based on the event rate of $10^3$ to $2\times 
10^{4}$ $\rm sky^{-1}\,day^{-1}$ as seen by Parkes \citep{Thornton13Sci}, the event rate for NS26m is 
from
$10^{-3}$ to $2\times 10^{-2}$ per day, i.e. 1 per 3 years to 1 per 50 days.  
The event rate at KM40m is rather tiny, and it falls in range of $[5\times 
10^{-5}, 10^{-4}]$ per day, i.e.  1 per 50 years to 1 per 3 years. The derived event rates are comparable using the method in \citet{Chawla2017apj}. Summing up, we 
expect one FRB on a monthly or yearly timescale for our current setups. Clearly, 
KM40m is not optimal for FRB searching due to the high temperature and narrow bandwidth, but we expect a better receiver frontend 
can significantly help the current situation. 

\section{Discovery of the intriguing peryton}
\label{sec:peryton}

At the time this paper is written, we have observed for about 1600 hours at NS26m and 
2400 hours at KM40m. So far no FRB has been found. However, an intriguing peryton 
was detected at NS26m.

\begin{figure}
	\includegraphics[width=\columnwidth]{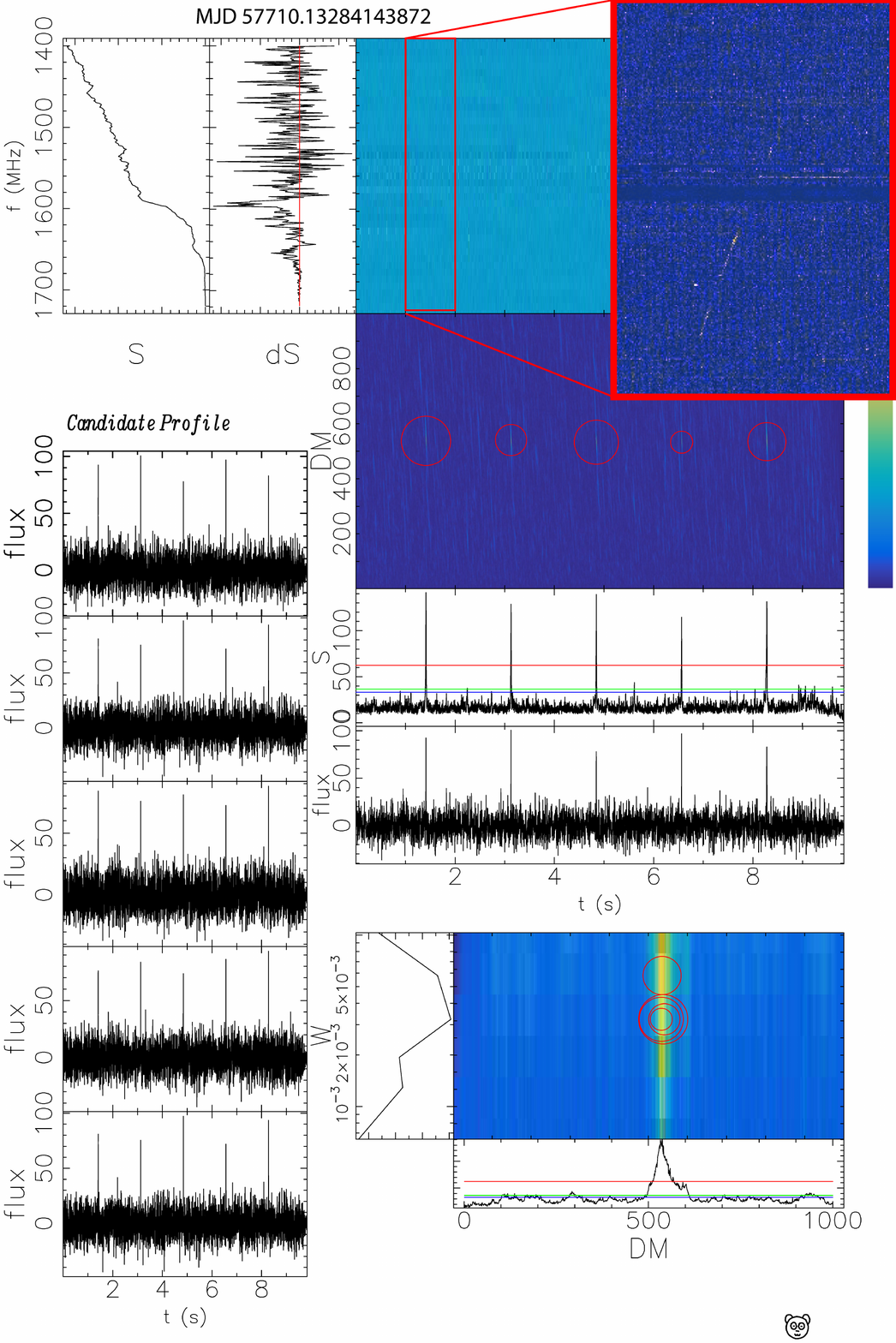}
		\caption{One of the peryton discovery plots generated by \BEA. 
		The meaning of each 
		panel is the same as \FIG{fig:pipeline_test}. The zoomed-in panel on the 
		top-right corner shows the details of a single burst. One can see the 
		dispersion feature rather clearly.
	}
    \label{fig:single_pulse}
\end{figure}

On November 18th 2016, between UTC 02:24 to 03:31 (local time 18th November, 08:24 to 
09:31), we detected a total of 218 broad-band radio pulses during pulsar timing observations, which show clear 
dispersion signature. The parameters of the pulsar timing observations are shown in \TAB{tab:timing_observations}. All the bursts have the same DM value of $531 \pm 5\, {\rm 
cm^{-3}\,pc}$. An example candidate plot generated by \BEA\ is shown in 
\FIG{fig:single_pulse}. The flux of each single burst is estimated as strong as 
20 Jy.

We found that the pulses spread across a 70-minute ``burst window''. 
The occurrences of pulses fall into three major timespans. 
Six pulses were recorded in the first timespan, when the telescope was slewing from PSR\,J1509+5531 to PSR\,J1239+2453. The pulses appeared sporadically. After the first 
burst window, no pulse occurred until the 47-th minute, where the second timespan 
starts, which lasted for 15 minutes. In this timespan, the telescope was pointed at PSR\,J1041$-$1942 and 145 pulses were recorded.
The third timespan started at the 55-th minute and lasted for 12 minutes, in which we recorded 67 pulses and the telescope was pointed at PSR\,J1012$-$2337.

\begin{table*}
	\centering
	\begin{threeparttable}
\caption{The parameters of the commensal pulsar timing observations, during which the peryton was detected.}

	\label{tab:timing_observations}
	\begin{tabular}{cccccc} 
		\hline\hline
		Pulsar & RAJ & DECJ & Start date & Start time & Observation\\
			    & (hh:mm:ss) & (dd:mm:ss) & (dd-mm-yy) & (UTC) & duration (s)\\
		\hline
		J1509+5531 & 15:09:25.6 & +55:31:32.4 & 2016-11-18 & 02:15:15 & 253\\
		J1239+2453 & 12:39:40.4 & +24:53:49.2 & 2016-11-18 & 02:26:25 & 367\\
		J0953+0755 & 09:53:09.3 & +07:55:35.8 & 2016-11-18 & 02:33:55 & 267\\
		J1041$-$1942 & 10:41:36.1 & -19:42:13.6 & 2016-11-18 & 02:39:15 & 975\\
		J1041$-$1942 & 10:41:36.1 & -19:42:13.6 & 2016-11-18 & 02:59:45 & 989\\
		J1012$-$2337 & 10:12:33.7 & -23:38:22.4 & 2016-11-18 & 03:16:35 & 989\\
		\hline
	\end{tabular}
\end{threeparttable}
	\end{table*}

\begin{figure}
	\includegraphics[width=\columnwidth]{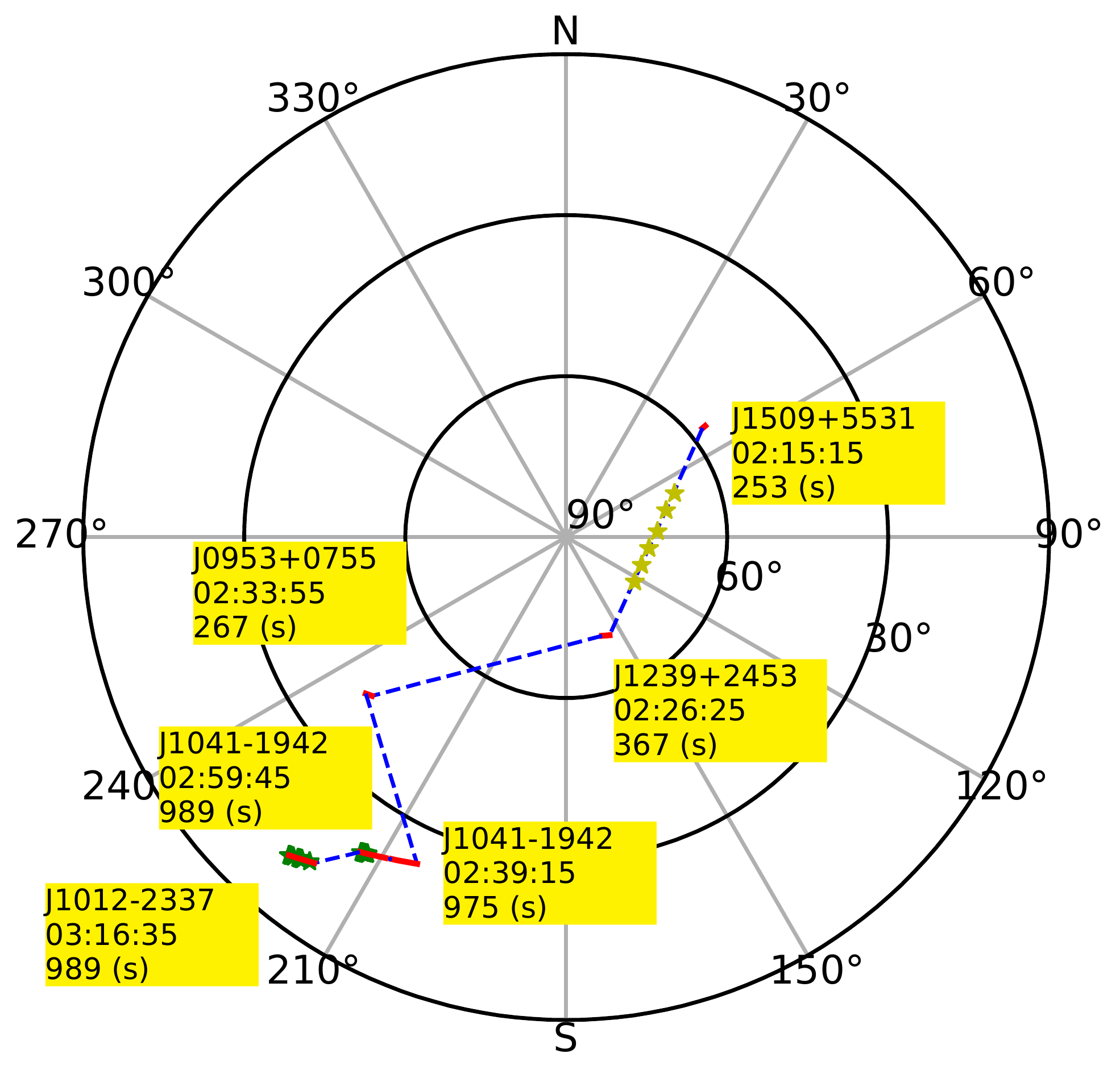}
		\caption{The orientation diagram with the occurrence altitude and azimuth angles of the 
		pulses detected at the Nanshan radio telescope. The thick red lines represent the duration of the pulsar timing observations with the parameters shown in \TAB{tab:timing_observations}. The blue dash lines represent the telescope slewing to the next pulsar. The six yellow stars represent the 6 pulses detected during the telescope slewing without exact locations and the green stars represent the pulses detected during the pulsar trackings. The locations of the total pulses are shown in \TAB{tab:bursts}.
We have also labelled the pulsar name, the start time and duration of each pulsar timing observation.}
    \label{fig:sky_regions}
\end{figure}

\begin{figure}
	\includegraphics[width=\columnwidth]{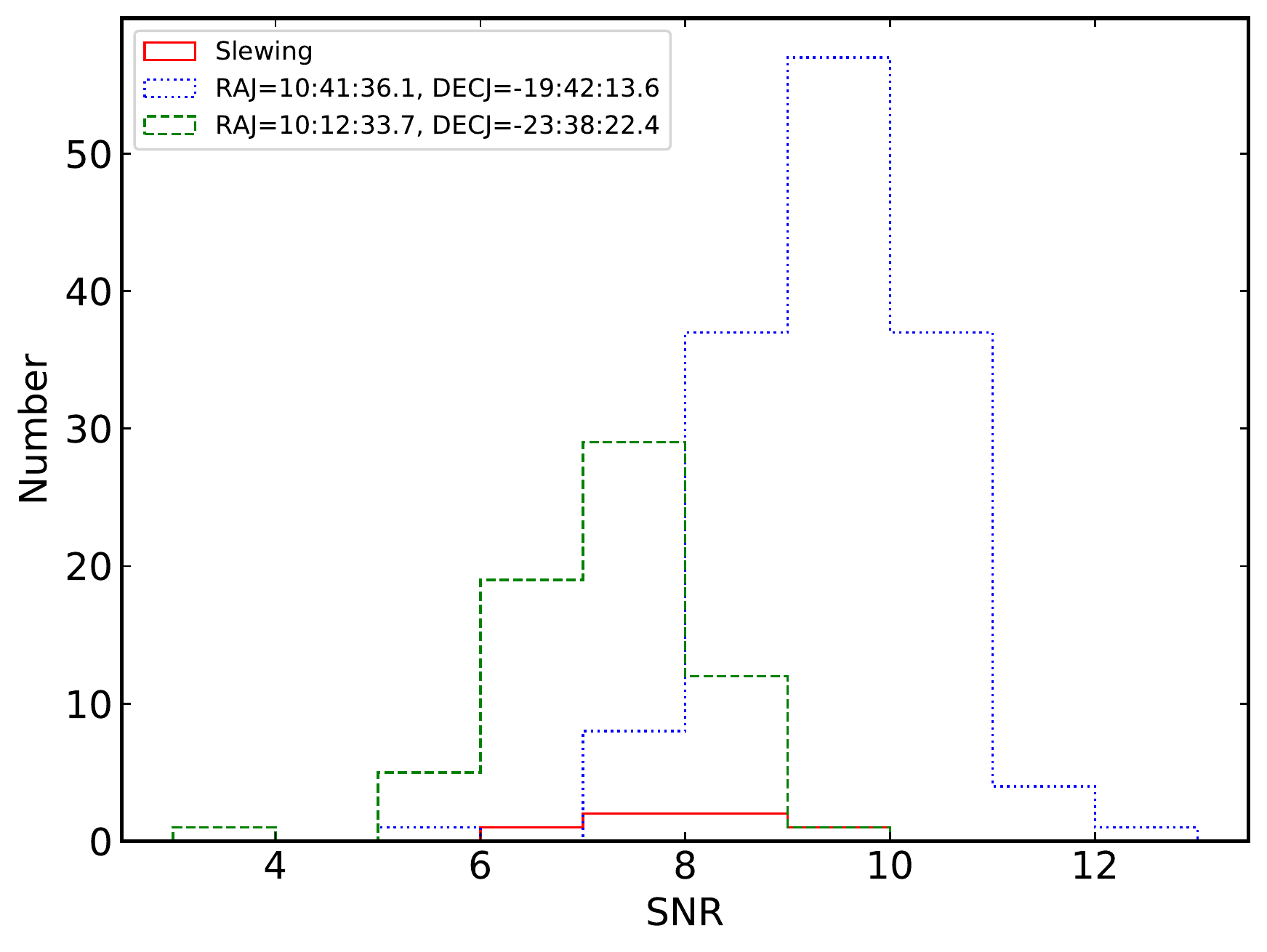}
		\caption{The histograms of the SNRs of the pulses in different sky locations.}
    \label{fig:snr}
\end{figure}

The burst positions and the telescope trackings are shown in \FIG{fig:sky_regions}. 
Since the telescope pointed in sky positions quite far apart
from each other
during the different detection timespans, one would expect
the SNR to be drastically different if the pulsed 
signal originated from a single celestial position. The distribution of the SNRs of the pulses in different sky regions are shown in \FIG{fig:snr}.
For a far-field source, we would expect more than $40$ dB variation between the 
pointing positions indicated in \FIG{fig:sky_regions}, estimated using the 
propagation model after taking telescope structure reflection into account 
\citep{has08}. We do not see such large signal amplitude variations,
which suggests that the pulsed signals do not come through the side lobes. The signal could potentially be understood in a scenario where it was picked up after the antenna feed, in which case the observed SNR would be roughly independent of the telescope pointing. However, this scenario requires an unlikely coincidence such that the signal is strong enough to overcome the approximately -90 dB isolation of coaxial cable and cavity for electronics and simultaneously not to be visible to the antenna feed. Otherwise we should detect a much stronger signal through the feed leakage. Thus, it appears more likely that the signal we report originates from local RFI.

\begin{figure}
	\includegraphics[width=\columnwidth]{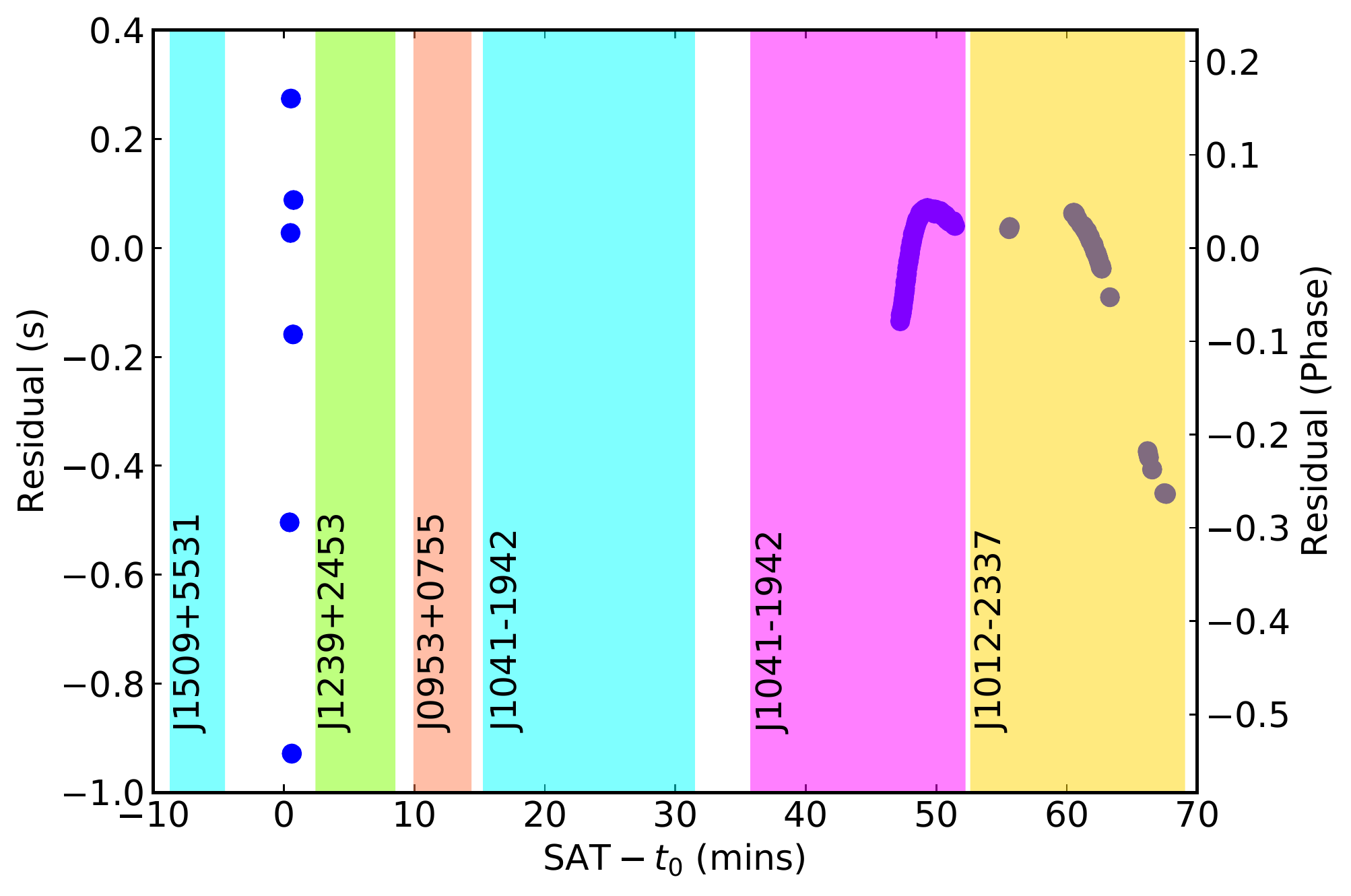}
		\caption{Timing residuals of the detected pulses. The x-axis shows the site arrival times (SATs) 
		in units of minute, where the reference epoch is $t_0={\rm MJD \,
		57710.1}$. 	
		The y-axis are the timing residuals in units of second and in units 
		of rotation phase.
		The shaded areas indicate the durations of pulsar timing observations.}
		\label{fig:timing}
\end{figure}

We nevertheless studied the timing behavior of the pulses. Following
the standard pulsar timing technique \citep[e. g.][]{HEM06}, 
we could measure and model the
times-of-arrival (TOAs) of the pulses. We aligned a few of the most bright
single pulses and then smoothed it to form the pulse profile template with
\textsc{psrchive} \citep{HVM04}. After measuring the TOA
of each single pulse, we used \textsc{tempo2} \citep{HEM06} to
build the timing model and calculate the timing residuals. 
The timing data indicate that the pulsed signal has a coherent timing behavior, where
we can measure the period to a rather good precision ($p=1.71287\pm 0.00004$ s).  
The post-fitting timing residuals are shown in \FIG{fig:timing}, where only the pulse
period is fitted. Since we detected no pulse between the first and the
second observing window, it is unclear if the timing solution
is still coherent for the first six data points. There seems to be a coherent
solution for the second and third observing window. If we further fit for 
the period derivative,
we get a value of $\dot{p}=3.50\times 10^{-6} \, {\rm s\, s^{-1}}$. From the
\FIG{fig:timing}, we can see that over the last 20 minutes, the residual
varies by 0.5 second.
The mechanism therefore which generates such a periodicity in
the pulses, must be stable in period to the $10^{-3}$ level.

\begin{figure}
	\includegraphics[width=\columnwidth]{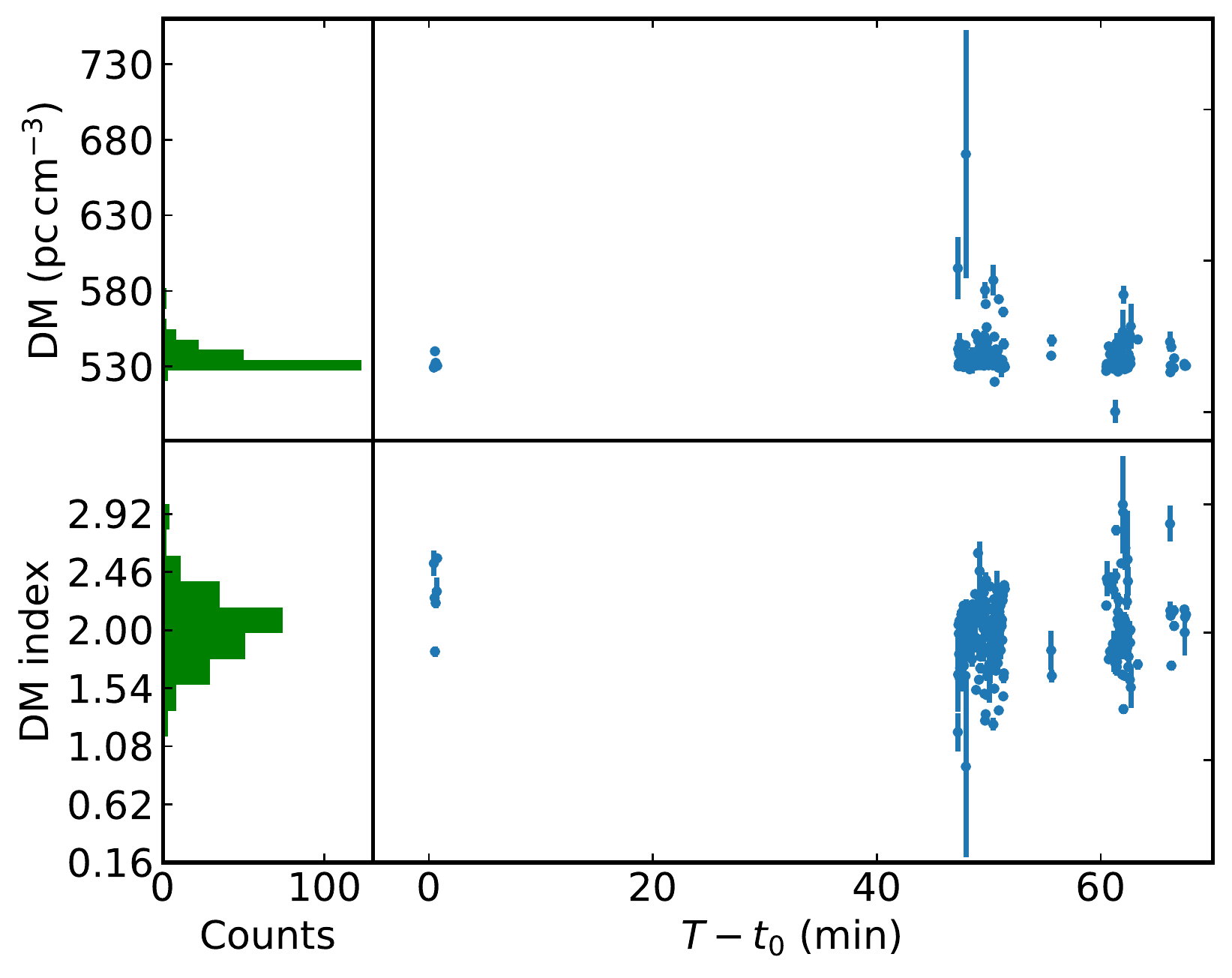}
		\caption{The DM and DM index for each single pulse. Top-left: The histogram 
		of DM values, the x-axis is the number of counts, and the y-axis is the 
		measured DM. Top-right: The time series of measures DM. The x-axis is the 
		time in minutes with reference epoch $t_0={\rm MJD \, 57710.1}$. Bottom-left, 
		the histogram of DM index. Bottom-right, the time series of measures DM 
		index.  As one can see, the pulses have similar DM values, 
		but a clear DM variation 
		is detected. A significant fraction ($\sim 20\%$) of pulses are compatible 
		with the cold plasma dispersion relation of DM index $\alpha= 2$.  }
		\label{fig:dmdmi}
\end{figure}

In order to investigate how the RFI signals disguise themselves as celestial
radio pulses, we measure the DM and dispersion index of each pulse, as shown in \TAB{tab:bursts}.
The dispersion index $\alpha$, as defined by the dispersive delay $\Delta
t\propto \nu{^{-\alpha}}$, will be 2 for radio wave propagating in
cold plasma \citep{Manchester80book}. 
Such an index is widely used to check if pulses are of
celestial origin \citep{Burke_Spolaor11apj, Petroff15}. 
We use a Bayesian approach to fit for both DM and
dispersion index simultaneously (Men et al., in prep). The measured
DM and dispersion index are shown in \FIG{fig:dmdmi}. As one can see,
the DM values cluster around the central value of $530\,{\rm cm^{-3}\,
pc}$. A clear variation of DM is also visible. The dispersion
indices of this peryton are also varying. Intriguingly,
the dispersion indices are around 2, and 17\% of pulses
have dispersion indices compatible with 2 within the 68\% error-bar.  The peryton 
would look like a true celestial source if only a small
fraction of the pulses was detected. To our knowledge, 
such a type of peryton has never been reported before.

\begin{figure}
	\includegraphics[width=\columnwidth]{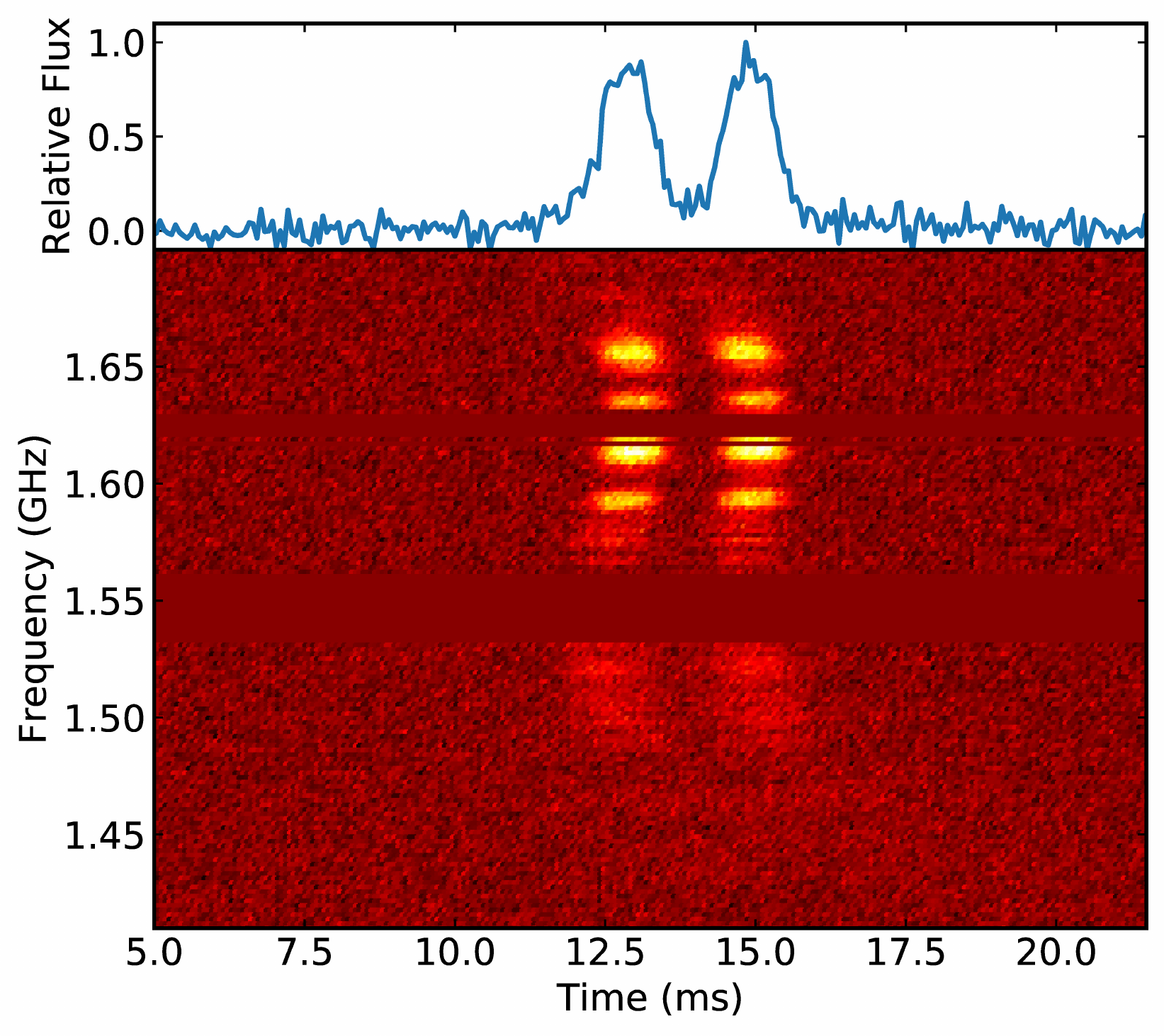}
		\caption{The perytons folded pulse. 
		Top panel: the de-dispersed and folded pulse 
		profile, where the x-axis is the time in units of millisecond
		and the y-axis is the 
		relative flux. We can see a clear double-peaked profile with 
		a total width of 4 ms.  
		Bottom panel: the de-dispersed pulse as function of frequency (y-axis) 
		and time (x-axis).  There is a residual curvature in the pulse, i.e. the 
		dispersion of the pulse does not follow the -2 index perfectly. The signal 
		does not come from a
		celestial origin. There is also frequency modulation across the full band, 
		which is visually similar to the scintillation behavior.  
		\label{fig:foldpulses}}
	\end{figure}

The peryton's spectrum structure can be made more clearly after summing up
individual pulses. The zoomed-in pulse is shown in \FIG{fig:foldpulses}. The 
time-integrated pulse is double-peaked, and shows scintillation-like structures 
in the spectrum. Thanks to the high SNR of the summed pulse, we can see that 
the pulse is not perfectly aligned across frequency after de-dispersing with 
$\alpha=2$, which also suggest that the signal dispersion does not originate 
from the celestial cold plasma. 

\section{Discussions and Summary}
\label{sec:dis}
This paper introduced the on-going FRB searching project using 
the Nanshan and Kunming radio telescopes. 
We described our searching hardware, software, data 
reduction algorithm and pipeline. We have
computed the expected detection sensitivity and event rate of Nanshan and 
Kunming radio telescopes. For Nanshan, the sensitivity and event rate are 1.1 Jy 
and $10^{-3}$ to $2\times 10^{-2}$ per day. For Kunming, the numbers are 3.3 Jy 
and $5\times 10^{-5}$ to $10^{-4}$ per day.
Based on the negative results in about $\rm 2400\ hrs \times \rm 0.04\  deg^2$ observations at S-band, we estimate a 95\% confidence upper limit on the FRB rate of  $R(S_\mathrm{min}=3.3\ \mathrm{Jy}) < 3.1\times10^4 \rm\ sky^{-1} day^{-1}$. This result agrees with the FRB rate reported in \citet{Burke-Spolaor2016apj}.

We introduced our data processing pipeline, \BEA, where we start with the 
likelihood ratio test and get the same filter as the matched-filter theory. We 
also use the equal-SNR-loss scheme to set up the most economic parameter
searching grid to save computational resources. As shown in \SEC{sec:rtpip}, the 
setup is a logarithmic grid in pulse width and linear grids in DM and pulse epoch.

To this point, we have not yet detected any FRBs, 
but we have detected and studied an intriguing peryton. 
The peryton detected at Nanshan 
radio telescope differs drastically 
from previously reported perytons \citep{Burke_Spolaor11apj, Petroff15} 
and has nearly identical properties compared to most of the currently known FRBs. 
Unlike the common peryton width of 300 ms, our peryton burst has 
a double-peaked pulse profile and a 2-ms width. The DM value (531 $\rm pc\,cm^{-3}$) 
of the burst is also similar to the currently known FRB population. The 
inspection of individual single pulse gives dispersion index close to -2. All 
these key features fall in the middle of the FRB parameter space. 

As shown in \FIG{fig:timing}, the peryton bursts 
lasted for a total of 70 minutes. Only six 
single bursts were detected in the first 45th minutes. In the rest of the observing 
window, i.e.  from the 45 minutes to the 70th minutes, we detected very regular 
pulses with a period of 1.71287 s. 
It is highly likely that the phase of the bursts 
is coherent.

Two major reasons lead us to suggest that the bursts we see are likely to be RFI. 
Firstly, we detected significant deviations of the DM index after adding 
the single bursts. Secondly, the telescope pointing has moved towards different 
directions in the detection window. We could not trace the origin of the 
peryton and it never showed up again in subsequent observations. 
We searched all the available data and we 
continue paying attention for similar signals in our FRB searching campaign. 
Up to the time of writing the current paper, we found no other similar perytons.  

With the available information, we did not conclude on a reasonable
explanation for the peryton signal. A category of perytons has
previously been identified to be generated from microwave ovens
\citep{Burke_Spolaor11apj, Petroff15}. Moreover, perytons with
quasi-periodicity of an approximate 22-s cycle were also previously reported 
\citep{Kocz2012}. However, the peryton
detected at Nanshan has remarkable different properties: 1) the
pulses occurred in a more precise period; 2) the widths of the
pulses are narrower, at $\sim$ 2 ms; 3) the DMs of the pulses are
more concentrated, at $\sim 531\ \rm pc\, cm^{-3}$, and the DM index is very close 
to $-2$ as given by the cold plasma dispersion. Therefore, the peryton detected 
at Nanshan is likely to be a different type. 
It does not seem very probable that the signal originates from a microwave oven. There are two major types of microwave ovens, the transformer type and the inverter type \citep{Matsumoto2003ec}. For most of the transformer type ovens, the pulse width modulation for the microwave-oven power control operates with period longer than 10 s. For the inverter type, the conversion frequency is around several kHz. To the best of the author's knowledge, periodic signal from microwave ovens similar to the peryton we discuss in this paper, was not previously reported \citep{Anderson1979nat, Yamanaka1995ec, Despres1997ec, Matsumoto2003ec}.
The signal is unlikely to originate 
from artificial satellite communication facilities
or airplane, otherwise we should see such signals quite often.
The lack of signal modulation also indicates that the burst may not
be communication signals. We also made sure the signal is not due
to instrumental instabilities or failure.  Since we were piggybacking
the pulsar observations, we can check if known pulsars are visible
in the data.  We found a single pulse with the correct DM of 9
$\rm{cm^{-3}\, pc}$ during observations of PSR J1239+2453
\citep{Kazantsev2017}. The origin of the Nanshan peryton therefore
remains unclear.

Our detected peryton mimics a real FRB signal. In fact, if only one or two 
single pulses were detected, one may had well concluded that the bursts are FRBs.
We found that the DM index is critical to evaluate whether the burst is of
celestial origin or not.

\FIG{fig:sky_regions} suggests that the apparent peryton positions on the sky 
may look like multiple isolated islands. Otherwise we would require the peryton 
to shut off from the 62th minute to the 65th minute as shown in 
\FIG{fig:timing}, while still keeping the coherent phase of timing. This suggests 
that the apparent directionality of near-field interference signals are rather 
complicated. For most of the FRB detection efforts, one relies on multibeam 
receivers to validate the celestial origin of burst signals. The idea is that
near-field RFI can appear in multiple beams, while far-field true FRB 
signal appears only in adjacent beams. The indication of a complicated near-field 
pattern suggests that we should be more careful and may need extra information to 
validate the celestial origin of the pulsed signals. 

\section*{Acknowledgments}

This work was supported by NSFC U15311243, National Basic Research Program
of China, 973 Program, 2015CB857101, XDB23010200, 11690024, 11373011, and 
funding from TianShanChuangXinTuanDui and the Max-Planck Partner Group.  We
 are grateful to R. N. Caballero for reading through the paper and giving helpful suggestions.


\bibliographystyle{mnras}
\bibliography{ms} 




%
%


\appendix
\section{The likelihood ratio test statistic}
\label{sec:appliktst}

The likelihood function is the probability distribution of data given the signal 
model. For the signal $\VEC{s}$ of pure Gaussian white noise, the likelihood is
\begin{equation}
	\Lambda_{\rm Null}=\frac{1}{(2 \upi \sigma^2)^{n/2}} 
	{\rm e}^{-\frac{\VEC{s}^2}{2\sigma^2}}\,,
	\label{eq:like1}
\end{equation}
where $n$ is the number of signal data points and $\sigma$ is the standard 
deviation of the noise.
When there is a square wave signal on top of the Gaussian noise, the 
likelihood is
\begin{equation}
	\Lambda_{\rm Null}=\frac{1}{(2 \upi \sigma^2)^{n/2}} 
	{\rm e}^{-\frac{\left(\VEC{s}-\VEC{h}\right)^2}{2\sigma^2}}\,,
	\label{eq:like2}
\end{equation}

The logarithm of the likelihood ratio between the cases of with and without a 
signal becomes
\begin{equation}
	S\equiv \log \left(\frac{\Lambda_{\rm Sig}}{\Lambda_{ \rm Null}}\right)\,,
	\label{eq:likelihoodratio}
\end{equation}
which leads to \EQ{eq:dets}.

\bsp	
\label{lastpage}

\onecolumn
\begin{longtable}{ccccccccccc}
\label{tab:bursts}\\
\caption{The properties of the bursts from the peryton detected at Nanshan.}\\
\hline
\hline
Burst & Date & Time & RAJ & DECJ & Telescope & Telescope & DM & DM index & SNR & Width\\
 & (dd-mm-yy) & (UTC) & (hh:mm:ss) & (dd:mm:ss) & altitude (deg) & azimuth (deg) & ($\rm pc\,cm^{-3}$) & & & (ms)\\
\hline
\endfirsthead
\multicolumn{11}{c}
{\tablename\ \thetable\ -- (\textit{Continued})}\\
\hline
\hline
Burst & Date & Time & RAJ & DECJ & Telescope & Telescope & DM & DM index & SNR & Width\\
 & (dd-mm-yy) & (UTC) & (hh:mm:ss) & (dd:mm:ss) & altitude (deg) & azimuth (deg) & ($\rm pc\,cm^{-3}$) & & & (ms)\\
\hline
\endhead
\endfoot
\hline
\endlastfoot
1 & 2016-11-18 & 02:24:26 & -- & -- & -- & -- & 529.3 $\pm$ 1.1 & 2.54 $\pm$ 0.10 & 6.4 & 2.0\\
2 & 2016-11-18 & 02:24:30 & -- & -- & -- & -- & 530.4 $\pm$ 1.1 & 2.27 $\pm$ 0.07 & 7.2 & 1.9\\
3 & 2016-11-18 & 02:24:32 & -- & -- & -- & -- & 540.1 $\pm$ 2.6 & 1.85 $\pm$ 0.04 & 9.2 & 1.9\\
4 & 2016-11-18 & 02:24:35 & -- & -- & -- & -- & 532.5 $\pm$ 0.5 & 2.23 $\pm$ 0.04 & 8.9 & 1.9\\
5 & 2016-11-18 & 02:24:41 & -- & -- & -- & -- & 530.9 $\pm$ 0.8 & 2.32 $\pm$ 0.11 & 9.3 & 1.9\\
6 & 2016-11-18 & 02:24:43 & -- & -- & -- & -- & 530.4 $\pm$ 0.8 & 2.58 $\pm$ 0.03 & 8.5 & 1.8\\
7 & 2016-11-18 & 03:11:13 & 10:41:36.1 & -19:42:13.6 & 20.305 & 212.085 & 595.0 $\pm$ 20.6 & 1.22 $\pm$ 0.15 & 10.5 & 1.3\\
8 & 2016-11-18 & 03:11:15 & 10:41:36.1 & -19:42:13.6 & 20.302 & 212.093 & 541.5 $\pm$ 5.6 & 1.67 $\pm$ 0.29 & 9.1 & 1.5\\
9 & 2016-11-18 & 03:11:17 & 10:41:36.1 & -19:42:13.6 & 20.299 & 212.100 & 530.2 $\pm$ 0.8 & 2.06 $\pm$ 0.05 & 9.5 & 1.5\\
10 & 2016-11-18 & 03:11:18 & 10:41:36.1 & -19:42:13.6 & 20.297 & 212.104 & 532.0 $\pm$ 0.9 & 1.99 $\pm$ 0.03 & 10.4 & 1.5\\
11 & 2016-11-18 & 03:11:20 & 10:41:36.1 & -19:42:13.6 & 20.294 & 212.112 & 538.7 $\pm$ 1.6 & 1.83 $\pm$ 0.03 & 11.8 & 1.4\\
12 & 2016-11-18 & 03:11:22 & 10:41:36.1 & -19:42:13.6 & 20.291 & 212.120 & 530.3 $\pm$ 1.1 & 2.09 $\pm$ 0.08 & 10.3 & 1.5\\
13 & 2016-11-18 & 03:11:24 & 10:41:36.1 & -19:42:13.6 & 20.287 & 212.127 & 545.5 $\pm$ 6.5 & 1.64 $\pm$ 0.08 & 9.5 & 1.6\\
14 & 2016-11-18 & 03:11:25 & 10:41:36.1 & -19:42:13.6 & 20.286 & 212.131 & 545.1 $\pm$ 1.8 & 1.72 $\pm$ 0.02 & 9.4 & 1.7\\
15 & 2016-11-18 & 03:11:27 & 10:41:36.1 & -19:42:13.6 & 20.283 & 212.139 & 532.7 $\pm$ 1.0 & 1.98 $\pm$ 0.03 & 10.5 & 1.3\\
16 & 2016-11-18 & 03:11:29 & 10:41:36.1 & -19:42:13.6 & 20.279 & 212.146 & 542.0 $\pm$ 5.2 & 1.72 $\pm$ 0.18 & 10.8 & 1.3\\
17 & 2016-11-18 & 03:11:30 & 10:41:36.1 & -19:42:13.6 & 20.278 & 212.150 & 534.5 $\pm$ 0.9 & 1.98 $\pm$ 0.03 & 9.7 & 1.6\\
18 & 2016-11-18 & 03:11:32 & 10:41:36.1 & -19:42:13.6 & 20.275 & 212.158 & 533.2 $\pm$ 0.6 & 2.14 $\pm$ 0.08 & 11.0 & 1.4\\
19 & 2016-11-18 & 03:11:34 & 10:41:36.1 & -19:42:13.6 & 20.271 & 212.165 & 537.9 $\pm$ 1.9 & 1.81 $\pm$ 0.03 & 10.3 & 1.5\\
20 & 2016-11-18 & 03:11:36 & 10:41:36.1 & -19:42:13.6 & 20.268 & 212.173 & 536.8 $\pm$ 1.7 & 1.83 $\pm$ 0.03 & 9.6 & 1.4\\
21 & 2016-11-18 & 03:11:37 & 10:41:36.1 & -19:42:13.6 & 20.266 & 212.177 & 530.8 $\pm$ 0.5 & 2.16 $\pm$ 0.03 & 9.8 & 1.4\\
22 & 2016-11-18 & 03:11:39 & 10:41:36.1 & -19:42:13.6 & 20.263 & 212.185 & 530.7 $\pm$ 0.9 & 2.00 $\pm$ 0.13 & 9.5 & 1.4\\
23 & 2016-11-18 & 03:11:41 & 10:41:36.1 & -19:42:13.6 & 20.260 & 212.192 & 530.8 $\pm$ 0.6 & 2.07 $\pm$ 0.04 & 10.1 & 1.4\\
24 & 2016-11-18 & 03:11:42 & 10:41:36.1 & -19:42:13.6 & 20.258 & 212.196 & 537.3 $\pm$ 1.8 & 1.79 $\pm$ 0.03 & 9.4 & 1.5\\
25 & 2016-11-18 & 03:11:44 & 10:41:36.1 & -19:42:13.6 & 20.255 & 212.204 & 529.6 $\pm$ 0.5 & 2.21 $\pm$ 0.03 & 10.4 & 1.8\\
26 & 2016-11-18 & 03:11:46 & 10:41:36.1 & -19:42:13.6 & 20.252 & 212.211 & 540.3 $\pm$ 6.2 & 1.74 $\pm$ 0.16 & 10.6 & 1.5\\
27 & 2016-11-18 & 03:11:48 & 10:41:36.1 & -19:42:13.6 & 20.249 & 212.219 & 533.5 $\pm$ 0.8 & 2.02 $\pm$ 0.03 & 10.8 & 1.7\\
28 & 2016-11-18 & 03:11:49 & 10:41:36.1 & -19:42:13.6 & 20.247 & 212.223 & 531.1 $\pm$ 0.6 & 2.08 $\pm$ 0.02 & 10.3 & 1.6\\
29 & 2016-11-18 & 03:11:51 & 10:41:36.1 & -19:42:13.6 & 20.244 & 212.230 & 534.7 $\pm$ 1.9 & 1.91 $\pm$ 0.04 & 10.2 & 1.4\\
30 & 2016-11-18 & 03:11:53 & 10:41:36.1 & -19:42:13.6 & 20.241 & 212.238 & 531.7 $\pm$ 1.2 & 1.95 $\pm$ 0.03 & 10.5 & 1.5\\
31 & 2016-11-18 & 03:11:54 & 10:41:36.1 & -19:42:13.6 & 20.239 & 212.242 & 544.1 $\pm$ 2.7 & 1.66 $\pm$ 0.03 & 9.8 & 1.7\\
32 & 2016-11-18 & 03:11:56 & 10:41:36.1 & -19:42:13.6 & 20.236 & 212.249 & 670.5 $\pm$ 81.9 & 0.95 $\pm$ 0.71 & 7.2 & 1.3\\
33 & 2016-11-18 & 03:11:58 & 10:41:36.1 & -19:42:13.6 & 20.233 & 212.257 & 531.0 $\pm$ 0.8 & 2.16 $\pm$ 0.10 & 10.9 & 1.5\\
34 & 2016-11-18 & 03:12:00 & 10:41:36.1 & -19:42:13.6 & 20.229 & 212.265 & 532.4 $\pm$ 1.2 & 1.95 $\pm$ 0.03 & 9.6 & 1.6\\
35 & 2016-11-18 & 03:12:01 & 10:41:36.1 & -19:42:13.6 & 20.228 & 212.269 & 531.3 $\pm$ 0.7 & 2.06 $\pm$ 0.02 & 10.1 & 1.4\\
36 & 2016-11-18 & 03:12:03 & 10:41:36.1 & -19:42:13.6 & 20.225 & 212.276 & 537.7 $\pm$ 1.5 & 1.82 $\pm$ 0.02 & 9.5 & 1.4\\
37 & 2016-11-18 & 03:12:05 & 10:41:36.1 & -19:42:13.6 & 20.221 & 212.284 & 534.9 $\pm$ 2.1 & 1.93 $\pm$ 0.04 & 10.1 & 1.8\\
38 & 2016-11-18 & 03:12:07 & 10:41:36.1 & -19:42:13.6 & 20.218 & 212.291 & 532.0 $\pm$ 1.4 & 2.04 $\pm$ 0.05 & 10.2 & 1.6\\
39 & 2016-11-18 & 03:12:08 & 10:41:36.1 & -19:42:13.6 & 20.216 & 212.295 & 530.7 $\pm$ 0.5 & 2.14 $\pm$ 0.05 & 10.5 & 1.4\\
40 & 2016-11-18 & 03:12:10 & 10:41:36.1 & -19:42:13.6 & 20.213 & 212.303 & 533.1 $\pm$ 0.9 & 2.11 $\pm$ 0.03 & 10.4 & 1.7\\
41 & 2016-11-18 & 03:12:12 & 10:41:36.1 & -19:42:13.6 & 20.210 & 212.311 & 534.0 $\pm$ 1.1 & 1.99 $\pm$ 0.09 & 9.7 & 1.5\\
42 & 2016-11-18 & 03:12:13 & 10:41:36.1 & -19:42:13.6 & 20.208 & 212.314 & 535.6 $\pm$ 1.9 & 1.86 $\pm$ 0.04 & 9.4 & 1.3\\
43 & 2016-11-18 & 03:12:15 & 10:41:36.1 & -19:42:13.6 & 20.205 & 212.322 & 533.6 $\pm$ 1.4 & 1.98 $\pm$ 0.09 & 9.1 & 1.5\\
44 & 2016-11-18 & 03:12:17 & 10:41:36.1 & -19:42:13.6 & 20.202 & 212.330 & 528.2 $\pm$ 0.6 & 2.12 $\pm$ 0.02 & 10.4 & 1.3\\
45 & 2016-11-18 & 03:12:19 & 10:41:36.1 & -19:42:13.6 & 20.199 & 212.337 & 530.1 $\pm$ 0.8 & 2.12 $\pm$ 0.03 & 10.1 & 1.4\\
46 & 2016-11-18 & 03:12:20 & 10:41:36.1 & -19:42:13.6 & 20.197 & 212.341 & 529.2 $\pm$ 0.7 & 2.06 $\pm$ 0.06 & 8.9 & 1.4\\
47 & 2016-11-18 & 03:12:22 & 10:41:36.1 & -19:42:13.6 & 20.194 & 212.349 & 531.2 $\pm$ 0.5 & 2.21 $\pm$ 0.03 & 9.6 & 1.6\\
48 & 2016-11-18 & 03:12:24 & 10:41:36.1 & -19:42:13.6 & 20.191 & 212.356 & 532.9 $\pm$ 0.7 & 2.15 $\pm$ 0.07 & 8.7 & 1.6\\
49 & 2016-11-18 & 03:12:25 & 10:41:36.1 & -19:42:13.6 & 20.189 & 212.360 & 531.4 $\pm$ 0.7 & 2.18 $\pm$ 0.03 & 9.4 & 1.6\\
50 & 2016-11-18 & 03:12:27 & 10:41:36.1 & -19:42:13.6 & 20.186 & 212.368 & 532.8 $\pm$ 1.3 & 1.93 $\pm$ 0.03 & 8.1 & 1.4\\
51 & 2016-11-18 & 03:12:29 & 10:41:36.1 & -19:42:13.6 & 20.182 & 212.375 & 538.5 $\pm$ 4.2 & 1.79 $\pm$ 0.06 & 10.5 & 1.6\\
52 & 2016-11-18 & 03:12:31 & 10:41:36.1 & -19:42:13.6 & 20.179 & 212.383 & 531.4 $\pm$ 6.1 & 2.04 $\pm$ 0.19 & 8.6 & 1.4\\
53 & 2016-11-18 & 03:12:32 & 10:41:36.1 & -19:42:13.6 & 20.178 & 212.387 & 531.1 $\pm$ 0.5 & 2.22 $\pm$ 0.03 & 8.7 & 1.5\\
54 & 2016-11-18 & 03:12:34 & 10:41:36.1 & -19:42:13.6 & 20.174 & 212.394 & 533.9 $\pm$ 1.7 & 1.89 $\pm$ 0.04 & 8.5 & 1.6\\
55 & 2016-11-18 & 03:12:37 & 10:41:36.1 & -19:42:13.6 & 20.170 & 212.406 & 533.1 $\pm$ 1.1 & 1.98 $\pm$ 0.08 & 9.4 & 1.4\\
56 & 2016-11-18 & 03:12:39 & 10:41:36.1 & -19:42:13.6 & 20.166 & 212.413 & 530.8 $\pm$ 1.2 & 2.11 $\pm$ 0.04 & 9.7 & 1.7\\
57 & 2016-11-18 & 03:12:41 & 10:41:36.1 & -19:42:13.6 & 20.163 & 212.421 & 533.3 $\pm$ 2.3 & 1.95 $\pm$ 0.08 & 8.7 & 1.4\\
58 & 2016-11-18 & 03:12:43 & 10:41:36.1 & -19:42:13.6 & 20.160 & 212.429 & 531.8 $\pm$ 0.7 & 2.12 $\pm$ 0.03 & 8.9 & 1.4\\
59 & 2016-11-18 & 03:12:44 & 10:41:36.1 & -19:42:13.6 & 20.158 & 212.433 & 530.5 $\pm$ 0.7 & 2.07 $\pm$ 0.02 & 9.6 & 1.4\\
60 & 2016-11-18 & 03:12:46 & 10:41:36.1 & -19:42:13.6 & 20.155 & 212.440 & 531.0 $\pm$ 0.5 & 2.30 $\pm$ 0.03 & 10.0 & 1.7\\
61 & 2016-11-18 & 03:12:48 & 10:41:36.1 & -19:42:13.6 & 20.152 & 212.448 & 532.0 $\pm$ 2.7 & 2.07 $\pm$ 0.13 & 8.8 & 1.7\\
62 & 2016-11-18 & 03:12:49 & 10:41:36.1 & -19:42:13.6 & 20.150 & 212.452 & 530.9 $\pm$ 0.5 & 2.17 $\pm$ 0.03 & 10.8 & 1.3\\
63 & 2016-11-18 & 03:12:51 & 10:41:36.1 & -19:42:13.6 & 20.147 & 212.459 & 551.1 $\pm$ 3.5 & 1.55 $\pm$ 0.03 & 8.1 & 1.6\\
64 & 2016-11-18 & 03:12:53 & 10:41:36.1 & -19:42:13.6 & 20.144 & 212.467 & 533.4 $\pm$ 3.1 & 1.89 $\pm$ 0.05 & 9.6 & 1.5\\
65 & 2016-11-18 & 03:12:55 & 10:41:36.1 & -19:42:13.6 & 20.140 & 212.474 & 533.6 $\pm$ 1.6 & 1.95 $\pm$ 0.04 & 8.6 & 1.5\\
66 & 2016-11-18 & 03:12:56 & 10:41:36.1 & -19:42:13.6 & 20.139 & 212.478 & 531.9 $\pm$ 0.8 & 2.12 $\pm$ 0.03 & 8.4 & 1.5\\
67 & 2016-11-18 & 03:12:58 & 10:41:36.1 & -19:42:13.6 & 20.135 & 212.486 & 531.0 $\pm$ 0.5 & 2.20 $\pm$ 0.03 & 10.6 & 1.5\\
68 & 2016-11-18 & 03:13:00 & 10:41:36.1 & -19:42:13.6 & 20.132 & 212.494 & 538.6 $\pm$ 1.8 & 1.91 $\pm$ 0.05 & 7.9 & 1.5\\
69 & 2016-11-18 & 03:13:01 & 10:41:36.1 & -19:42:13.6 & 20.131 & 212.497 & 547.2 $\pm$ 1.9 & 2.62 $\pm$ 0.04 & 8.3 & 1.4\\
70 & 2016-11-18 & 03:13:03 & 10:41:36.1 & -19:42:13.6 & 20.127 & 212.505 & 531.0 $\pm$ 0.5 & 2.17 $\pm$ 0.06 & 9.0 & 1.3\\
71 & 2016-11-18 & 03:13:05 & 10:41:36.1 & -19:42:13.6 & 20.124 & 212.513 & 534.9 $\pm$ 2.9 & 1.89 $\pm$ 0.06 & 9.1 & 1.6\\
72 & 2016-11-18 & 03:13:07 & 10:41:36.1 & -19:42:13.6 & 20.121 & 212.520 & 547.4 $\pm$ 3.5 & 1.63 $\pm$ 0.03 & 8.6 & 1.3\\
73 & 2016-11-18 & 03:13:08 & 10:41:36.1 & -19:42:13.6 & 20.119 & 212.524 & 533.6 $\pm$ 1.7 & 2.27 $\pm$ 0.11 & 8.7 & 1.4\\
74 & 2016-11-18 & 03:13:10 & 10:41:36.1 & -19:42:13.6 & 20.116 & 212.532 & 542.6 $\pm$ 5.2 & 2.48 $\pm$ 0.23 & 7.4 & 1.6\\
75 & 2016-11-18 & 03:13:12 & 10:41:36.1 & -19:42:13.6 & 20.113 & 212.539 & 531.0 $\pm$ 1.0 & 2.20 $\pm$ 0.05 & 9.8 & 1.5\\
76 & 2016-11-18 & 03:13:13 & 10:41:36.1 & -19:42:13.6 & 20.111 & 212.543 & 540.1 $\pm$ 2.8 & 1.72 $\pm$ 0.04 & 8.5 & 1.4\\
77 & 2016-11-18 & 03:13:15 & 10:41:36.1 & -19:42:13.6 & 20.108 & 212.551 & 533.6 $\pm$ 1.4 & 1.81 $\pm$ 0.02 & 8.8 & 1.4\\
78 & 2016-11-18 & 03:13:17 & 10:41:36.1 & -19:42:13.6 & 20.105 & 212.558 & 532.2 $\pm$ 0.8 & 2.14 $\pm$ 0.03 & 8.3 & 1.6\\
79 & 2016-11-18 & 03:13:19 & 10:41:36.1 & -19:42:13.6 & 20.101 & 212.566 & 531.0 $\pm$ 0.5 & 2.18 $\pm$ 0.03 & 9.7 & 1.4\\
80 & 2016-11-18 & 03:13:20 & 10:41:36.1 & -19:42:13.6 & 20.100 & 212.570 & 532.8 $\pm$ 0.8 & 2.09 $\pm$ 0.06 & 8.9 & 1.4\\
81 & 2016-11-18 & 03:13:22 & 10:41:36.1 & -19:42:13.6 & 20.096 & 212.577 & 534.8 $\pm$ 1.0 & 2.07 $\pm$ 0.04 & 10.0 & 1.5\\
82 & 2016-11-18 & 03:13:24 & 10:41:36.1 & -19:42:13.6 & 20.093 & 212.585 & 536.4 $\pm$ 0.8 & 2.20 $\pm$ 0.05 & 7.7 & 1.5\\
83 & 2016-11-18 & 03:13:25 & 10:41:36.1 & -19:42:13.6 & 20.092 & 212.589 & 535.2 $\pm$ 1.5 & 1.95 $\pm$ 0.03 & 9.4 & 1.5\\
84 & 2016-11-18 & 03:13:27 & 10:41:36.1 & -19:42:13.6 & 20.088 & 212.596 & 533.8 $\pm$ 2.9 & 1.84 $\pm$ 0.05 & 8.7 & 1.5\\
85 & 2016-11-18 & 03:13:29 & 10:41:36.1 & -19:42:13.6 & 20.085 & 212.604 & 537.2 $\pm$ 2.5 & 1.87 $\pm$ 0.10 & 9.4 & 1.5\\
86 & 2016-11-18 & 03:13:30 & 10:41:36.1 & -19:42:13.6 & 20.083 & 212.608 & 531.8 $\pm$ 1.2 & 2.03 $\pm$ 0.05 & 9.1 & 1.5\\
87 & 2016-11-18 & 03:13:32 & 10:41:36.1 & -19:42:13.6 & 20.080 & 212.615 & 535.9 $\pm$ 1.5 & 2.31 $\pm$ 0.11 & 8.8 & 1.4\\
88 & 2016-11-18 & 03:13:34 & 10:41:36.1 & -19:42:13.6 & 20.077 & 212.623 & 530.6 $\pm$ 1.6 & 2.10 $\pm$ 0.08 & 8.8 & 1.4\\
89 & 2016-11-18 & 03:13:36 & 10:41:36.1 & -19:42:13.6 & 20.074 & 212.631 & 550.2 $\pm$ 3.4 & 1.52 $\pm$ 0.03 & 8.5 & 1.4\\
90 & 2016-11-18 & 03:13:37 & 10:41:36.1 & -19:42:13.6 & 20.072 & 212.634 & 531.8 $\pm$ 0.9 & 2.15 $\pm$ 0.07 & 9.2 & 1.5\\
91 & 2016-11-18 & 03:13:39 & 10:41:36.1 & -19:42:13.6 & 20.069 & 212.642 & 580.5 $\pm$ 5.3 & 1.31 $\pm$ 0.03 & 8.7 & 1.5\\
92 & 2016-11-18 & 03:13:41 & 10:41:36.1 & -19:42:13.6 & 20.065 & 212.650 & 533.9 $\pm$ 0.5 & 2.24 $\pm$ 0.03 & 9.2 & 1.5\\
93 & 2016-11-18 & 03:13:42 & 10:41:36.1 & -19:42:13.6 & 20.064 & 212.653 & 571.3 $\pm$ 0.8 & 1.36 $\pm$ 0.00 & 8.5 & 1.6\\
94 & 2016-11-18 & 03:13:44 & 10:41:36.1 & -19:42:13.6 & 20.061 & 212.661 & 540.1 $\pm$ 0.9 & 2.41 $\pm$ 0.06 & 9.0 & 1.6\\
95 & 2016-11-18 & 03:13:46 & 10:41:36.1 & -19:42:13.6 & 20.057 & 212.669 & 532.1 $\pm$ 0.8 & 2.01 $\pm$ 0.02 & 9.5 & 1.3\\
96 & 2016-11-18 & 03:13:48 & 10:41:36.1 & -19:42:13.6 & 20.054 & 212.676 & 556.0 $\pm$ 3.2 & 1.51 $\pm$ 0.03 & 8.6 & 1.4\\
97 & 2016-11-18 & 03:13:49 & 10:41:36.1 & -19:42:13.6 & 20.052 & 212.680 & 539.6 $\pm$ 2.6 & 1.71 $\pm$ 0.03 & 9.4 & 1.4\\
98 & 2016-11-18 & 03:13:51 & 10:41:36.1 & -19:42:13.6 & 20.049 & 212.688 & 546.1 $\pm$ 3.6 & 1.66 $\pm$ 0.04 & 8.8 & 1.7\\
99 & 2016-11-18 & 03:13:53 & 10:41:36.1 & -19:42:13.6 & 20.046 & 212.695 & 533.2 $\pm$ 1.2 & 2.03 $\pm$ 0.06 & 9.8 & 1.6\\
100 & 2016-11-18 & 03:13:54 & 10:41:36.1 & -19:42:13.6 & 20.044 & 212.699 & 539.1 $\pm$ 2.8 & 1.71 $\pm$ 0.04 & 9.1 & 1.4\\
101 & 2016-11-18 & 03:13:56 & 10:41:36.1 & -19:42:13.6 & 20.041 & 212.707 & 541.5 $\pm$ 2.7 & 1.72 $\pm$ 0.03 & 9.7 & 1.6\\
102 & 2016-11-18 & 03:13:58 & 10:41:36.1 & -19:42:13.6 & 20.038 & 212.714 & 530.9 $\pm$ 0.9 & 2.18 $\pm$ 0.04 & 9.8 & 1.7\\
103 & 2016-11-18 & 03:14:00 & 10:41:36.1 & -19:42:13.6 & 20.034 & 212.722 & 531.3 $\pm$ 0.5 & 2.19 $\pm$ 0.03 & 9.2 & 1.4\\
104 & 2016-11-18 & 03:14:01 & 10:41:36.1 & -19:42:13.6 & 20.033 & 212.726 & 536.9 $\pm$ 2.1 & 1.75 $\pm$ 0.30 & 8.6 & 1.4\\
105 & 2016-11-18 & 03:14:03 & 10:41:36.1 & -19:42:13.6 & 20.029 & 212.733 & 539.0 $\pm$ 2.1 & 1.72 $\pm$ 0.03 & 9.7 & 1.2\\
106 & 2016-11-18 & 03:14:05 & 10:41:36.1 & -19:42:13.6 & 20.026 & 212.741 & 539.1 $\pm$ 0.3 & 2.36 $\pm$ 0.02 & 8.8 & 1.4\\
107 & 2016-11-18 & 03:14:06 & 10:41:36.1 & -19:42:13.6 & 20.025 & 212.745 & 536.3 $\pm$ 4.8 & 1.83 $\pm$ 0.23 & 7.5 & 1.4\\
108 & 2016-11-18 & 03:14:08 & 10:41:36.1 & -19:42:13.6 & 20.021 & 212.752 & 538.4 $\pm$ 3.7 & 1.75 $\pm$ 0.11 & 8.5 & 1.6\\
109 & 2016-11-18 & 03:14:10 & 10:41:36.1 & -19:42:13.6 & 20.018 & 212.760 & 532.3 $\pm$ 1.4 & 1.90 $\pm$ 0.03 & 8.4 & 1.4\\
110 & 2016-11-18 & 03:14:12 & 10:41:36.1 & -19:42:13.6 & 20.015 & 212.768 & 535.3 $\pm$ 1.7 & 1.88 $\pm$ 0.04 & 9.4 & 1.6\\
111 & 2016-11-18 & 03:14:13 & 10:41:36.1 & -19:42:13.6 & 20.013 & 212.771 & 537.4 $\pm$ 3.2 & 1.82 $\pm$ 0.07 & 9.8 & 1.4\\
112 & 2016-11-18 & 03:14:15 & 10:41:36.1 & -19:42:13.6 & 20.010 & 212.779 & 532.4 $\pm$ 1.0 & 2.05 $\pm$ 0.03 & 10.8 & 1.5\\
113 & 2016-11-18 & 03:14:17 & 10:41:36.1 & -19:42:13.6 & 20.007 & 212.787 & 531.0 $\pm$ 0.6 & 2.10 $\pm$ 0.03 & 11.6 & 1.3\\
114 & 2016-11-18 & 03:14:18 & 10:41:36.1 & -19:42:13.6 & 20.005 & 212.790 & 532.2 $\pm$ 0.5 & 2.18 $\pm$ 0.03 & 10.3 & 1.6\\
115 & 2016-11-18 & 03:14:20 & 10:41:36.1 & -19:42:13.6 & 20.002 & 212.798 & 530.8 $\pm$ 1.5 & 1.98 $\pm$ 0.04 & 7.9 & 1.4\\
116 & 2016-11-18 & 03:14:22 & 10:41:36.1 & -19:42:13.6 & 19.998 & 212.806 & 531.2 $\pm$ 2.2 & 1.95 $\pm$ 0.11 & 8.8 & 1.5\\
117 & 2016-11-18 & 03:14:24 & 10:41:36.1 & -19:42:13.6 & 19.995 & 212.813 & 587.1 $\pm$ 10.0 & 1.28 $\pm$ 0.05 & 8.7 & 1.4\\
118 & 2016-11-18 & 03:14:25 & 10:41:36.1 & -19:42:13.6 & 19.993 & 212.817 & 533.1 $\pm$ 2.1 & 2.11 $\pm$ 0.12 & 9.6 & 1.6\\
119 & 2016-11-18 & 03:14:27 & 10:41:36.1 & -19:42:13.6 & 19.990 & 212.825 & 533.1 $\pm$ 0.5 & 2.26 $\pm$ 0.03 & 10.3 & 1.6\\
120 & 2016-11-18 & 03:14:29 & 10:41:36.1 & -19:42:13.6 & 19.987 & 212.832 & 549.7 $\pm$ 3.0 & 1.56 $\pm$ 0.03 & 9.9 & 1.5\\
121 & 2016-11-18 & 03:14:30 & 10:41:36.1 & -19:42:13.6 & 19.985 & 212.836 & 519.9 $\pm$ 0.2 & 2.00 $\pm$ 0.00 & 7.7 & 6.5\\
122 & 2016-11-18 & 03:14:32 & 10:41:36.1 & -19:42:13.6 & 19.982 & 212.844 & 533.5 $\pm$ 1.8 & 1.92 $\pm$ 0.07 & 9.5 & 1.6\\
123 & 2016-11-18 & 03:14:34 & 10:41:36.1 & -19:42:13.6 & 19.979 & 212.851 & 530.8 $\pm$ 1.0 & 2.04 $\pm$ 0.03 & 10.2 & 1.5\\
124 & 2016-11-18 & 03:14:36 & 10:41:36.1 & -19:42:13.6 & 19.975 & 212.859 & 530.9 $\pm$ 1.4 & 1.97 $\pm$ 0.03 & 8.9 & 1.5\\
125 & 2016-11-18 & 03:14:37 & 10:41:36.1 & -19:42:13.6 & 19.974 & 212.862 & 541.4 $\pm$ 2.1 & 1.70 $\pm$ 0.03 & 10.0 & 1.5\\
126 & 2016-11-18 & 03:14:39 & 10:41:36.1 & -19:42:13.6 & 19.970 & 212.870 & 531.6 $\pm$ 0.8 & 2.06 $\pm$ 0.03 & 10.7 & 1.6\\
127 & 2016-11-18 & 03:14:41 & 10:41:36.1 & -19:42:13.6 & 19.967 & 212.878 & 538.0 $\pm$ 3.6 & 1.78 $\pm$ 0.09 & 9.4 & 1.4\\
128 & 2016-11-18 & 03:14:42 & 10:41:36.1 & -19:42:13.6 & 19.966 & 212.881 & 532.9 $\pm$ 1.2 & 2.33 $\pm$ 0.15 & 9.1 & 1.5\\
129 & 2016-11-18 & 03:14:44 & 10:41:36.1 & -19:42:13.6 & 19.962 & 212.889 & 533.3 $\pm$ 1.4 & 1.90 $\pm$ 0.04 & 9.6 & 1.5\\
130 & 2016-11-18 & 03:14:46 & 10:41:36.1 & -19:42:13.6 & 19.959 & 212.897 & 533.3 $\pm$ 2.1 & 1.93 $\pm$ 0.05 & 9.3 & 1.4\\
131 & 2016-11-18 & 03:14:48 & 10:41:36.1 & -19:42:13.6 & 19.956 & 212.904 & 540.6 $\pm$ 2.8 & 1.76 $\pm$ 0.06 & 9.3 & 1.4\\
132 & 2016-11-18 & 03:14:49 & 10:41:36.1 & -19:42:13.6 & 19.954 & 212.908 & 531.1 $\pm$ 1.4 & 2.07 $\pm$ 0.04 & 10.9 & 1.7\\
133 & 2016-11-18 & 03:14:51 & 10:41:36.1 & -19:42:13.6 & 19.951 & 212.916 & 529.2 $\pm$ 0.5 & 2.24 $\pm$ 0.05 & 10.6 & 1.7\\
134 & 2016-11-18 & 03:14:53 & 10:41:36.1 & -19:42:13.6 & 19.947 & 212.923 & 574.5 $\pm$ 2.5 & 1.39 $\pm$ 0.01 & 9.4 & 1.7\\
135 & 2016-11-18 & 03:14:56 & 10:41:36.1 & -19:42:13.6 & 19.943 & 212.935 & 529.3 $\pm$ 0.7 & 2.16 $\pm$ 0.03 & 11.0 & 1.7\\
136 & 2016-11-18 & 03:14:58 & 10:41:36.1 & -19:42:13.6 & 19.939 & 212.942 & 532.0 $\pm$ 1.0 & 2.02 $\pm$ 0.03 & 10.8 & 1.5\\
137 & 2016-11-18 & 03:15:00 & 10:41:36.1 & -19:42:13.6 & 19.936 & 212.950 & 531.1 $\pm$ 0.7 & 2.30 $\pm$ 0.07 & 10.5 & 1.8\\
138 & 2016-11-18 & 03:15:01 & 10:41:36.1 & -19:42:13.6 & 19.934 & 212.954 & 529.0 $\pm$ 0.5 & 2.21 $\pm$ 0.03 & 9.7 & 1.8\\
139 & 2016-11-18 & 03:15:03 & 10:41:36.1 & -19:42:13.6 & 19.931 & 212.961 & 534.3 $\pm$ 2.7 & 1.86 $\pm$ 0.07 & 8.6 & 1.6\\
140 & 2016-11-18 & 03:15:05 & 10:41:36.1 & -19:42:13.6 & 19.928 & 212.969 & 530.8 $\pm$ 0.4 & 2.23 $\pm$ 0.03 & 10.6 & 1.5\\
141 & 2016-11-18 & 03:15:06 & 10:41:36.1 & -19:42:13.6 & 19.926 & 212.973 & 529.2 $\pm$ 6.5 & 2.07 $\pm$ 0.18 & 9.9 & 1.5\\
142 & 2016-11-18 & 03:15:08 & 10:41:36.1 & -19:42:13.6 & 19.923 & 212.980 & 530.0 $\pm$ 4.6 & 2.05 $\pm$ 0.17 & 9.6 & 1.4\\
143 & 2016-11-18 & 03:15:10 & 10:41:36.1 & -19:42:13.6 & 19.919 & 212.988 & 531.4 $\pm$ 0.8 & 2.10 $\pm$ 0.04 & 12.0 & 1.6\\
144 & 2016-11-18 & 03:15:12 & 10:41:36.1 & -19:42:13.6 & 19.916 & 212.995 & 534.4 $\pm$ 1.3 & 1.94 $\pm$ 0.03 & 9.4 & 1.7\\
145 & 2016-11-18 & 03:15:13 & 10:41:36.1 & -19:42:13.6 & 19.915 & 212.999 & 528.9 $\pm$ 0.4 & 2.25 $\pm$ 0.03 & 9.7 & 1.6\\
146 & 2016-11-18 & 03:15:15 & 10:41:36.1 & -19:42:13.6 & 19.911 & 213.007 & 530.8 $\pm$ 0.6 & 2.29 $\pm$ 0.05 & 7.6 & 1.7\\
147 & 2016-11-18 & 03:15:17 & 10:41:36.1 & -19:42:13.6 & 19.908 & 213.014 & 566.2 $\pm$ 3.5 & 1.50 $\pm$ 0.03 & 9.6 & 1.8\\
148 & 2016-11-18 & 03:15:18 & 10:41:36.1 & -19:42:13.6 & 19.906 & 213.018 & 545.0 $\pm$ 3.9 & 1.65 $\pm$ 0.05 & 8.4 & 1.3\\
149 & 2016-11-18 & 03:15:20 & 10:41:36.1 & -19:42:13.6 & 19.903 & 213.026 & 544.6 $\pm$ 2.6 & 1.68 $\pm$ 0.03 & 8.8 & 1.7\\
150 & 2016-11-18 & 03:15:22 & 10:41:36.1 & -19:42:13.6 & 19.900 & 213.033 & 530.9 $\pm$ 0.5 & 2.37 $\pm$ 0.03 & 5.8 & 1.6\\
151 & 2016-11-18 & 03:15:25 & 10:41:36.1 & -19:42:13.6 & 19.895 & 213.045 & 529.6 $\pm$ 0.5 & 2.34 $\pm$ 0.04 & 9.6 & 1.7\\
152 & 2016-11-18 & 03:19:34 & 10:12:33.7 & -23:38:22.4 & 12.900 & 218.329 & 537.2 $\pm$ 1.7 & 1.86 $\pm$ 0.15 & 7.1 & 1.7\\
153 & 2016-11-18 & 03:19:37 & 10:12:33.7 & -23:38:22.4 & 12.895 & 218.339 & 547.3 $\pm$ 3.8 & 1.66 $\pm$ 0.05 & 7.5 & 1.3\\
154 & 2016-11-18 & 03:24:28 & 10:12:33.7 & -23:38:22.4 & 12.344 & 219.329 & 527.1 $\pm$ 0.6 & 2.21 $\pm$ 0.03 & 6.3 & 1.6\\
155 & 2016-11-18 & 03:24:30 & 10:12:33.7 & -23:38:22.4 & 12.341 & 219.336 & 529.9 $\pm$ 0.9 & 2.21 $\pm$ 0.04 & 6.0 & 1.7\\
156 & 2016-11-18 & 03:24:32 & 10:12:33.7 & -23:38:22.4 & 12.337 & 219.342 & 531.8 $\pm$ 2.1 & 2.42 $\pm$ 0.14 & 7.4 & 0.7\\
157 & 2016-11-18 & 03:24:37 & 10:12:33.7 & -23:38:22.4 & 12.327 & 219.359 & 531.6 $\pm$ 0.7 & 2.39 $\pm$ 0.05 & 7.7 & 2.1\\
158 & 2016-11-18 & 03:24:42 & 10:12:33.7 & -23:38:22.4 & 12.318 & 219.376 & 543.3 $\pm$ 2.3 & 1.79 $\pm$ 0.03 & 6.8 & 1.4\\
159 & 2016-11-18 & 03:24:45 & 10:12:33.7 & -23:38:22.4 & 12.312 & 219.386 & 530.4 $\pm$ 0.9 & 2.41 $\pm$ 0.06 & 6.2 & 1.6\\
160 & 2016-11-18 & 03:24:49 & 10:12:33.7 & -23:38:22.4 & 12.304 & 219.400 & 538.1 $\pm$ 2.4 & 1.85 $\pm$ 0.04 & 8.3 & 1.6\\
161 & 2016-11-18 & 03:25:04 & 10:12:33.7 & -23:38:22.4 & 12.276 & 219.451 & 535.5 $\pm$ 2.2 & 1.91 $\pm$ 0.04 & 7.2 & 1.4\\
162 & 2016-11-18 & 03:25:08 & 10:12:33.7 & -23:38:22.4 & 12.268 & 219.464 & 528.5 $\pm$ 0.8 & 2.33 $\pm$ 0.07 & 5.7 & 1.9\\
163 & 2016-11-18 & 03:25:09 & 10:12:33.7 & -23:38:22.4 & 12.266 & 219.468 & 539.6 $\pm$ 6.3 & 1.85 $\pm$ 0.16 & 8.4 & 1.6\\
164 & 2016-11-18 & 03:25:16 & 10:12:33.7 & -23:38:22.4 & 12.253 & 219.491 & 500.2 $\pm$ 7.6 & 1.76 $\pm$ 0.09 & 3.9 & 3.5\\
165 & 2016-11-18 & 03:25:18 & 10:12:33.7 & -23:38:22.4 & 12.249 & 219.498 & 529.8 $\pm$ 0.7 & 2.44 $\pm$ 0.06 & 9.8 & 1.5\\
166 & 2016-11-18 & 03:25:21 & 10:12:33.7 & -23:38:22.4 & 12.243 & 219.508 & 542.2 $\pm$ 1.6 & 2.80 $\pm$ 0.04 & 7.2 & 1.8\\
167 & 2016-11-18 & 03:25:25 & 10:12:33.7 & -23:38:22.4 & 12.235 & 219.522 & 545.6 $\pm$ 1.9 & 1.73 $\pm$ 0.02 & 7.5 & 1.6\\
168 & 2016-11-18 & 03:25:26 & 10:12:33.7 & -23:38:22.4 & 12.233 & 219.525 & 542.7 $\pm$ 9.7 & 1.77 $\pm$ 0.11 & 7.9 & 1.8\\
169 & 2016-11-18 & 03:25:28 & 10:12:33.7 & -23:38:22.4 & 12.230 & 219.532 & 532.8 $\pm$ 1.7 & 2.10 $\pm$ 0.21 & 6.9 & 1.4\\
170 & 2016-11-18 & 03:25:30 & 10:12:33.7 & -23:38:22.4 & 12.226 & 219.539 & 537.2 $\pm$ 2.7 & 1.85 $\pm$ 0.05 & 7.3 & 1.6\\
171 & 2016-11-18 & 03:25:32 & 10:12:33.7 & -23:38:22.4 & 12.222 & 219.545 & 526.6 $\pm$ 1.6 & 2.16 $\pm$ 0.07 & 6.4 & 1.6\\
172 & 2016-11-18 & 03:25:33 & 10:12:33.7 & -23:38:22.4 & 12.220 & 219.549 & 529.2 $\pm$ 4.0 & 2.06 $\pm$ 0.08 & 7.8 & 1.7\\
173 & 2016-11-18 & 03:25:35 & 10:12:33.7 & -23:38:22.4 & 12.216 & 219.555 & 528.0 $\pm$ 0.6 & 2.25 $\pm$ 0.05 & 6.7 & 1.7\\
174 & 2016-11-18 & 03:25:37 & 10:12:33.7 & -23:38:22.4 & 12.212 & 219.562 & 534.5 $\pm$ 2.7 & 1.98 $\pm$ 0.08 & 7.9 & 1.5\\
175 & 2016-11-18 & 03:25:38 & 10:12:33.7 & -23:38:22.4 & 12.210 & 219.566 & 531.6 $\pm$ 1.6 & 2.00 $\pm$ 0.04 & 8.3 & 1.6\\
176 & 2016-11-18 & 03:25:40 & 10:12:33.7 & -23:38:22.4 & 12.206 & 219.572 & 531.4 $\pm$ 0.9 & 2.06 $\pm$ 0.12 & 7.7 & 1.7\\
177 & 2016-11-18 & 03:25:44 & 10:12:33.7 & -23:38:22.4 & 12.199 & 219.586 & 533.5 $\pm$ 1.7 & 1.94 $\pm$ 0.04 & 8.9 & 1.8\\
178 & 2016-11-18 & 03:25:45 & 10:12:33.7 & -23:38:22.4 & 12.197 & 219.589 & 535.8 $\pm$ 3.8 & 1.86 $\pm$ 0.07 & 6.9 & 1.8\\
179 & 2016-11-18 & 03:25:47 & 10:12:33.7 & -23:38:22.4 & 12.193 & 219.596 & 531.5 $\pm$ 2.5 & 2.03 $\pm$ 0.09 & 6.9 & 1.5\\
180 & 2016-11-18 & 03:25:49 & 10:12:33.7 & -23:38:22.4 & 12.189 & 219.603 & 530.4 $\pm$ 0.7 & 2.54 $\pm$ 0.03 & 8.1 & 1.6\\
181 & 2016-11-18 & 03:25:50 & 10:12:33.7 & -23:38:22.4 & 12.187 & 219.606 & 528.7 $\pm$ 1.3 & 2.10 $\pm$ 0.06 & 7.5 & 1.6\\
182 & 2016-11-18 & 03:25:54 & 10:12:33.7 & -23:38:22.4 & 12.180 & 219.620 & 550.8 $\pm$ 2.7 & 1.67 $\pm$ 0.03 & 6.1 & 1.6\\
183 & 2016-11-18 & 03:25:55 & 10:12:33.7 & -23:38:22.4 & 12.178 & 219.623 & 538.0 $\pm$ 1.6 & 1.85 $\pm$ 0.03 & 7.9 & 1.2\\
184 & 2016-11-18 & 03:25:57 & 10:12:33.7 & -23:38:22.4 & 12.174 & 219.630 & 553.0 $\pm$ 14.5 & 3.00 $\pm$ 0.38 & 5.5 & 1.1\\
185 & 2016-11-18 & 03:25:59 & 10:12:33.7 & -23:38:22.4 & 12.170 & 219.636 & 546.9 $\pm$ 1.6 & 2.94 $\pm$ 0.03 & 5.5 & 1.5\\
186 & 2016-11-18 & 03:26:01 & 10:12:33.7 & -23:38:22.4 & 12.166 & 219.643 & 577.5 $\pm$ 5.8 & 1.40 $\pm$ 0.04 & 6.2 & 1.5\\
187 & 2016-11-18 & 03:26:02 & 10:12:33.7 & -23:38:22.4 & 12.164 & 219.647 & 535.9 $\pm$ 2.2 & 1.92 $\pm$ 0.05 & 8.7 & 1.5\\
188 & 2016-11-18 & 03:26:04 & 10:12:33.7 & -23:38:22.4 & 12.160 & 219.653 & 534.6 $\pm$ 1.4 & 1.91 $\pm$ 0.03 & 8.0 & 1.8\\
189 & 2016-11-18 & 03:26:06 & 10:12:33.7 & -23:38:22.4 & 12.157 & 219.660 & 548.9 $\pm$ 2.9 & 1.66 $\pm$ 0.03 & 8.1 & 1.8\\
190 & 2016-11-18 & 03:26:07 & 10:12:33.7 & -23:38:22.4 & 12.155 & 219.663 & 528.2 $\pm$ 2.7 & 2.10 $\pm$ 0.06 & 5.8 & 1.3\\
191 & 2016-11-18 & 03:26:09 & 10:12:33.7 & -23:38:22.4 & 12.151 & 219.670 & 536.4 $\pm$ 4.3 & 1.86 $\pm$ 0.07 & 8.4 & 2.0\\
192 & 2016-11-18 & 03:26:11 & 10:12:33.7 & -23:38:22.4 & 12.147 & 219.677 & 535.2 $\pm$ 3.6 & 2.66 $\pm$ 0.17 & 7.8 & 1.9\\
193 & 2016-11-18 & 03:26:13 & 10:12:33.7 & -23:38:22.4 & 12.143 & 219.684 & 533.8 $\pm$ 1.5 & 1.98 $\pm$ 0.16 & 7.2 & 1.3\\
194 & 2016-11-18 & 03:26:14 & 10:12:33.7 & -23:38:22.4 & 12.141 & 219.687 & 528.6 $\pm$ 1.0 & 2.07 $\pm$ 0.03 & 8.6 & 1.5\\
195 & 2016-11-18 & 03:26:16 & 10:12:33.7 & -23:38:22.4 & 12.137 & 219.694 & 533.4 $\pm$ 1.2 & 1.99 $\pm$ 0.04 & 7.9 & 2.1\\
196 & 2016-11-18 & 03:26:18 & 10:12:33.7 & -23:38:22.4 & 12.133 & 219.701 & 532.7 $\pm$ 1.6 & 2.00 $\pm$ 0.04 & 7.9 & 1.8\\
197 & 2016-11-18 & 03:26:19 & 10:12:33.7 & -23:38:22.4 & 12.131 & 219.704 & 530.6 $\pm$ 0.7 & 2.24 $\pm$ 0.06 & 7.4 & 1.5\\
198 & 2016-11-18 & 03:26:21 & 10:12:33.7 & -23:38:22.4 & 12.128 & 219.711 & 535.9 $\pm$ 2.1 & 1.88 $\pm$ 0.04 & 7.5 & 1.7\\
199 & 2016-11-18 & 03:26:23 & 10:12:33.7 & -23:38:22.4 & 12.124 & 219.717 & 532.2 $\pm$ 3.9 & 2.57 $\pm$ 0.38 & 6.3 & 1.4\\
200 & 2016-11-18 & 03:26:25 & 10:12:33.7 & -23:38:22.4 & 12.120 & 219.724 & 529.1 $\pm$ 2.3 & 2.40 $\pm$ 0.11 & 8.0 & 1.6\\
201 & 2016-11-18 & 03:26:26 & 10:12:33.7 & -23:38:22.4 & 12.118 & 219.728 & 543.8 $\pm$ 2.1 & 1.73 $\pm$ 0.03 & 6.4 & 1.6\\
202 & 2016-11-18 & 03:26:28 & 10:12:33.7 & -23:38:22.4 & 12.114 & 219.734 & 538.2 $\pm$ 3.5 & 1.81 $\pm$ 0.06 & 6.9 & 1.9\\
203 & 2016-11-18 & 03:26:33 & 10:12:33.7 & -23:38:22.4 & 12.104 & 219.751 & 533.5 $\pm$ 1.2 & 2.01 $\pm$ 0.08 & 7.6 & 1.3\\
204 & 2016-11-18 & 03:26:35 & 10:12:33.7 & -23:38:22.4 & 12.101 & 219.758 & 550.7 $\pm$ 3.6 & 1.63 $\pm$ 0.04 & 7.1 & 1.9\\
205 & 2016-11-18 & 03:26:37 & 10:12:33.7 & -23:38:22.4 & 12.097 & 219.765 & 535.1 $\pm$ 1.4 & 1.92 $\pm$ 0.03 & 7.9 & 1.1\\
206 & 2016-11-18 & 03:26:38 & 10:12:33.7 & -23:38:22.4 & 12.095 & 219.768 & 531.8 $\pm$ 1.0 & 2.02 $\pm$ 0.03 & 7.5 & 1.6\\
207 & 2016-11-18 & 03:26:40 & 10:12:33.7 & -23:38:22.4 & 12.091 & 219.775 & 556.6 $\pm$ 14.9 & 1.57 $\pm$ 0.16 & 5.6 & 1.5\\
208 & 2016-11-18 & 03:27:18 & 10:12:33.7 & -23:38:22.4 & 12.018 & 219.903 & 547.9 $\pm$ 3.0 & 1.75 $\pm$ 0.04 & 8.1 & 1.9\\
209 & 2016-11-18 & 03:30:10 & 10:12:33.7 & -23:38:22.4 & 11.683 & 220.480 & 546.2 $\pm$ 6.7 & 2.85 $\pm$ 0.14 & 6.2 & 1.6\\
210 & 2016-11-18 & 03:30:12 & 10:12:33.7 & -23:38:22.4 & 11.679 & 220.487 & 526.2 $\pm$ 0.9 & 2.17 $\pm$ 0.07 & 7.9 & 2.0\\
211 & 2016-11-18 & 03:30:14 & 10:12:33.7 & -23:38:22.4 & 11.675 & 220.494 & 530.6 $\pm$ 0.8 & 2.13 $\pm$ 0.03 & 6.4 & 1.6\\
212 & 2016-11-18 & 03:30:17 & 10:12:33.7 & -23:38:22.4 & 11.669 & 220.504 & 542.9 $\pm$ 3.3 & 1.74 $\pm$ 0.04 & 7.8 & 1.7\\
213 & 2016-11-18 & 03:30:31 & 10:12:33.7 & -23:38:22.4 & 11.642 & 220.550 & 529.1 $\pm$ 0.6 & 2.17 $\pm$ 0.04 & 7.7 & 1.6\\
214 & 2016-11-18 & 03:30:33 & 10:12:33.7 & -23:38:22.4 & 11.638 & 220.557 & 535.4 $\pm$ 1.3 & 2.05 $\pm$ 0.04 & 7.8 & 1.7\\
215 & 2016-11-18 & 03:31:27 & 10:12:33.7 & -23:38:22.4 & 11.532 & 220.738 & 531.8 $\pm$ 0.6 & 2.18 $\pm$ 0.03 & 6.5 & 1.7\\
216 & 2016-11-18 & 03:31:29 & 10:12:33.7 & -23:38:22.4 & 11.528 & 220.744 & 530.5 $\pm$ 1.3 & 2.00 $\pm$ 0.18 & 6.9 & 1.6\\
217 & 2016-11-18 & 03:31:31 & 10:12:33.7 & -23:38:22.4 & 11.524 & 220.751 & 530.7 $\pm$ 0.8 & 2.12 $\pm$ 0.04 & 7.6 & 1.7\\
218 & 2016-11-18 & 03:31:36 & 10:12:33.7 & -23:38:22.4 & 11.514 & 220.768 & 530.5 $\pm$ 0.9 & 2.14 $\pm$ 0.05 & 6.9 & 1.6\\
\end{longtable}
\end{document}